\documentclass[]{aastex631}

\shortauthors{Ng et al.}
\usepackage[T1]{fontenc}%
\usepackage{bm}
\usepackage{tabularx}
\usepackage{booktabs}
\usepackage{longtable,array,threeparttablex}
\usepackage{threeparttable}

\begin{document}

\title{Unveiling Mass Transfer in Solar Flares: Insights from Elemental Abundance Evolutions Observed by Chang'E-2 Solar X-ray Monitor}

\correspondingauthor{Xiaoping Zhang}
\email{xpzhang@must.edu.mo}

\author[0000-0001-6206-0138]{Man-Hei Ng}
\altaffiliation{These authors contributed equally to this work.}
\affiliation{State Key Laboratory of Lunar and Planetary Sciences, Macau University of Science and Technology, Macau 999078, China}
\affiliation{China National Space Administration (CNSA), Macau Center for Space Exploration and Science, Macau 999078, China}

\author[0009-0008-8352-6281]{Chi-Long Tang}
\altaffiliation{These authors contributed equally to this work.}
\affiliation{State Key Laboratory of Lunar and Planetary Sciences, Macau University of Science and Technology, Macau 999078, China}
\affiliation{China National Space Administration (CNSA), Macau Center for Space Exploration and Science, Macau 999078, China}

\author[0000-0002-4306-5213]{Xiaoping Zhang}
\affiliation{State Key Laboratory of Lunar and Planetary Sciences, Macau University of Science and Technology, Macau 999078, China}
\affiliation{China National Space Administration (CNSA), Macau Center for Space Exploration and Science, Macau 999078, China}

\author[0000-0002-0834-0187]{Kuan-Vai Tam}
\affiliation{School of Astronomy and Space Science, Nanjing University, Nanjing 210023, China}

\author[0000-0002-7289-642X]{Peng-Fei Chen}
\affiliation{School of Astronomy and Space Science, Nanjing University, Nanjing 210023, China}

\author{Wudong Dong}
\affiliation{Guangdong Planning and Designing Institute of Telecommunications Co., Ltd., Guangzhou 510630, China}

\author{Jing Li}
\affiliation{Department of Earth, Planetary and Space Sciences, University of California at Los Angeles, Los Angeles, CA 90095-1567, USA}

\author[0000-0001-6367-9530]{Chi-Pui Tang}
\affiliation{State Key Laboratory of Lunar and Planetary Sciences, Macau University of Science and Technology, Macau 999078, China}
\affiliation{China National Space Administration (CNSA), Macau Center for Space Exploration and Science, Macau 999078, China}

\begin{abstract}

Understanding how elemental abundances evolve during solar flares helps shed light on the mass and energy transfer between different solar atmospheric layers.
However, prior studies have mostly concentrated on averaged abundances or specific flare phases, leaving a gap in exploring the comprehensive observations throughout the entire flare process.
Consequently, investigations into this area are relatively scarce.
Exploiting the Solar X-ray Monitor data obtained from the Chang'E-2 lunar orbiter, we present two comprehensive soft X-ray spectroscopic observations of flares in active regions, AR 11149 and 11158, demonstrating elemental abundance evolutions under different conditions.
Our findings unveil the inverse first ionization potential (IFIP) effect during flares for Fe for the first time, and reaffirm its existence for Si.
Additionally, we observed a rare depletion of elemental abundances, marking the second IFIP effect in flare decay phases.
Our study offers a CSHKP model-based interpretation to elucidate the formation of both the FIP and IFIP effects in flare dynamics, with the inertia effect being incorporated into the ponderomotive force fractionation model.

\end{abstract}


\section{Introduction}
\label{sec_intro}

Solar flares are among the most explosive phenomena in the solar atmosphere, releasing enormous energy, ranging from $10^{29}$ to $10^{32}$~erg \citep{shibataSolarFlaresMagnetohydrodynamic2011, aulanierStandardFlareModel2013}.
The release of energy during a solar flare, powered by magnetic reconnection, efficiently accelerates electrons, leading to sudden and rapid bursts of electromagnetic radiation (see \citealp{benzFlareObservations2008}, \citealp{shibataSolarFlaresMagnetohydrodynamic2011}, and \citealp{chenCoronalMassEjections2011} for reviews).

In the standard CSHKP flare model \citep{carmichaelProcessFlares1964, sturrockModelHighEnergyPhase1966, hirayamaTheoreticalModelFlares1974, koppMagneticReconnectionCorona1976}, magnetic reconnection occurs in a current sheet, where the coronal magnetic energy is converted to kinetic, thermal, and non-thermal energies.
A large amount of particles are accelerated to near-relativistic speeds.
Some of them escape from the reconnection site and travel down to bombard the flare loop top and further down to the dense chromosphere.
Note that some electrons are accelerated by the termination shock at the flare loop top \citep{chenCoronalMassEjections2011}.
The electron beam loses all its energy and momentum in the dense plasma owing to Coulomb collisions, and creates a region of high temperature and pressure. This process is accompanied by hard X-ray (HXR) emissions at the flare loop top and footpoints through non-thermal Bremsstrahlung \citep[e.g.,][]{linNonthermalProcessesLarge1976}.
The high pressure pushes the heated plasma upward with a velocity of hundreds of km~s$^{-1}$ to fill the flare loop, leading to the generation of soft X-ray (SXR) radiation through thermal bremsstrahlung.
This is the well-known chromospheric evaporation \citep[e.g.,][]{neupertComparisonSolarXRay1968, fisherFlareLoopRadiative1985, forbesFormationFlareLoops1989} which explains the temporal evolution of all the thermal and non-thermal emissions (see reviews by \citealp{priestMagneticNatureSolar2002}, \citealp{langSunEarthSky2006}, and \citealp{benzFlareObservations2008}).

In the 1960s, it was discovered that the elemental abundances in the corona differ from those in the underlying photosphere, i.e., the elements with low first ionization potential (FIP) are enriched in the corona.
As reviewed by \citet{lamingFIPInverseFIP2015}, low-FIP ($\lesssim$10~eV) elements such as Ca, Mg, Fe, and Si are typically enhanced by a factor of $\sim$1.5--4 in the solar corona compared to those in the photosphere, while high-FIP ($>$10~eV) elements such as Ar, O, and S retain similar abundances to those in the photosphere \citep[e.g.,][]{meyerSolarstellarOuterAtmospheres1985, feldmanElementalAbundancesUpper1992, sabaSpectroscopicMeasurementsElement1995, fludraAbsoluteCoronalAbundances1999, feldmanElementAbundancesUpper2000, phillipsSOLARFLAREIRON2012, schmelzCOMPOSITIONSOLARCORONA2012, narendranathElementalAbundancesSolar2014, narendranathCoronalElementalAbundance2020, warrenMEASUREMENTSABSOLUTEABUNDANCES2014, dennisSOLARFLAREELEMENT2015, mondalEvolutionElementalAbundances2021, toEvolutionPlasmaComposition2021, namaCoronalElementalAbundances2023}.
This phenomenon was known as the ``FIP effect''. Such a feature was later revealed in solar flares as well, and its study plays an important role in understanding the physical mechanisms responsible for the mass and energy transfer between different layers of the solar atmosphere during solar flares \citep{bakerTransientInverseFIPPlasma2019, bakerCanSubphotosphericMagnetic2020, katsudaInverseFirstIonization2020, toEvolutionPlasmaComposition2021, mondalEvolutionElementalAbundances2021}.
On the other hand, it was sometimes found that the low-FIP elements are depleted in the corona than in the photosphere, which was termed ``inverse FIP (IFIP) effect''.
The IFIP effect is generally observed in highly complex solar active regions (ARs) as recently reported by \citet{doschekAnomalousRelativeAr2015}, \citet{doschekMYSTERIOUSCASESOLAR2016, doschekSunspotsStarspotsElemental2017}, and \citet{bakerTransientInverseFIPPlasma2019, bakerCanSubphotosphericMagnetic2020}.

For example, \citet{doschekMYSTERIOUSCASESOLAR2016, doschekSunspotsStarspotsElemental2017} investigated the behavior of the FIP effect in several ARs with relatively large sunspots. Their observations with the Hinode \citep{kosugiHinodeSolarBMission2007}/Extreme-ultraviolet Imaging Spectrometer \citep[EIS;][]{culhaneEUVImagingSpectrometer2007} suggested a suppression of the FIP effect near sunspots.
\citet{bakerTransientInverseFIPPlasma2019, bakerCanSubphotosphericMagnetic2020} observed an enhancement of the \ion{Ar}{14}/\ion{Ca}{14} intensity ratio at the footpoints of loops during the decay phase of two confined M-class flares (M2.2 and M1.4) in AR 11429, and one X-class flare (X9.3) in AR 12673 with Hinode/EIS.
The patches with enhanced \ion{Ar}{14}/\ion{Ca}{14} intensity ratios, indicative of the IFIP effect, were observed 10 minutes after the flare peak and disappeared $\sim$40 minutes later.
These observations suggested that the preferential locations for observing IFIP effect plasma are in the umbrae of coalescing sunspots.
The generation of the IFIP effect was thought to result from the subchromospheric or subphotospheric reconnection during sunspot coalescence, which causes wave refraction from below the chromosphere.
This interpretation was based on the ponderomotive force fractionation model \citep{lamingUnifiedPictureFirst2004, lamingTHERMALCONDUCTIVITYELEMENT2009, lamingNONWKBMODELSFIRST2009, lamingNONWKBMODELSFIRST2012, lamingFIPInverseFIP2015, lamingFirstIonizationPotential2017, lamingElementAbundancesNew2019}.
\citet{katsudaInverseFirstIonization2020} determined the elemental abundances of Ca, Si, S, and Ar (relative to H) for four giant X-class flares (X17.0, X5.4, X6.2, and X9.0) observed by the Suzaku \citep{koyamaXRayImagingSpectrometer2007}/X-ray Imaging Spectrometer \citep[XIS;][]{mitsudaXRayObservatorySuzaku2007}. The depletions in Si and S abundances were revealed at or around the peaks of these flares, with averages of $\sim$0.7 and $\sim$0.3 times their solar photospheric abundances, respectively, indicating the presence of the IFIP effect.
It should be noted that some observations of solar flares have also shown low-FIP element values that are close to photospheric levels. For instance, a mean FIP bias (defined as the ratio of the measured coronal elemental abundance with respect to the photospheric one) of 1.17 has been reported for Fe \citep{warrenMEASUREMENTSABSOLUTEABUNDANCES2014}, suggesting that the bulk of the flare plasma is evaporated from deeper regions within the chromosphere, below the layer where fractionation takes place, a point also implied by \citet{dennisSOLARFLAREELEMENT2015}.

The ponderomotive force fractionation model \citep{lamingUnifiedPictureFirst2004, lamingTHERMALCONDUCTIVITYELEMENT2009, lamingNONWKBMODELSFIRST2009, lamingNONWKBMODELSFIRST2012, lamingFIPInverseFIP2015, lamingFirstIonizationPotential2017, lamingElementAbundancesNew2019}, as known as Laming's model, offers a unified picture of the FIP and IFIP effects, providing insights into the physical processes responsible for the observed evolution of elemental abundances.
In Laming's model, FIP fractionation occurs as a result of the reflection of downward propagating coronal Alfvén waves in the chromosphere, mainly in regions with high temperature and density gradients.
The upward propagating Alfvén waves generate an upward ponderomotive force acting only on the ions rather than the neutrals.
Under typical solar conditions, the chromosphere favors the ionization of low-FIP elements whereas most high-FIP elements remain in neutral states.
With the upward ponderomotive force, a significant amount of low-FIP elements are pushed up, causing the enhancement of the low-FIP elements in the corona, which is the FIP effect.
On the other hand, in sunspot regions, a downward ponderomotive force is anticipated.
This force occurs when upward propagating $p$-mode waves originating from below the chromosphere are probably converted to fast-mode waves at the equipartition layer (i.e., where the Alfvén speed equals the sound speed), subsequently leading to the downward refraction of these up-going fast-mode waves in the chromosphere. This downward force causes the depletion of low-FIP elements in the plasma.

More interestingly, although many flares possess the FIP effect similar to the quiet corona, some flares exhibit distinct features.
Measurements conducted by the Solar Assembly for X-rays \citep[SAX;][]{schlemmXRaySpectrometerMESSENGER2007} of the MESSENGER mission \citep{goldMESSENGERMissionMercury2001, santoMESSENGERMissionMercury2001, solomonMESSENGERMissionMercury2001} and the data analyzed by \citet{dennisSOLARFLAREELEMENT2015} revealed differences between the flare abundances in their studies \citep{dennisSOLARFLAREELEMENT2015} and those in previous research on coronal abundances \citep{feldmanElementalAbundancesUpper1992}, hybrid abundances \citep{fludraAbsoluteCoronalAbundances1999}, and flare abundances \citep{warrenMEASUREMENTSABSOLUTEABUNDANCES2014, sylwesterSolarFlareComposition2014, phillipsSOLARFLAREIRON2012}.
In particular, \citet{dennisSOLARFLAREELEMENT2015} found significant deviations in the averaged abundances of Fe and Si from flare to flare.
Comparative analysis with their respective photospheric abundances unveiled average enhancements of 1.66-fold for Fe and 1.64-fold for Si across numerous flare events \citep{dennisSOLARFLAREELEMENT2015, asplundChemicalCompositionSun2009}.
In the corona, however, the Fe and Si elements exhibited enhancements of 3.98-fold and 3.89-fold, respectively, relative to their photospheric abundances \citep{feldmanElementalAbundancesUpper1992, asplundChemicalCompositionSun2009}.
These findings, in line with Laming's model on the FIP effect for coronal plasmas, led \citet{dennisSOLARFLAREELEMENT2015} to propose that the fractionation in their flare sample may occur for elements with FIPs of less than $\sim$7~eV, rather than $\sim$10~eV.
It can be seen that studying elemental abundances during flares is of utmost importance in enhancing the understanding of these phenomena and the associated underlying mechanisms.

In addition, the elemental abundances evolving during flares \citep[e.g.,][]{narendranathCoronalElementalAbundance2020, mondalEvolutionElementalAbundances2021, namaCoronalElementalAbundances2023} could be intimately related to the transfer and dissipation of the energy needed to heat the plasma to coronal temperatures \citep{mondalEvolutionElementalAbundances2021}.
Understanding the physical mechanisms responsible for this could help solve the coronal heating problem (see \citealp{klimchukSolvingCoronalHeating2006, klimchukNanoflareHeatingObservations2017} for reviews), which refers to the challenge of explaining how the corona attains temperatures two orders of magnitude hotter than the solar surface \citep[e.g.,][]{parkerTopologicalDissipationSmallScale1972, cargillNanoflareHeatingCorona2004, cargillActiveRegionEmission2014}, or at least help refine coronal heating models through the solar SXR spectral analysis \citep[e.g.,][]{winebargerDEFININGBLINDSPOT2012, caspiNEWOBSERVATIONSSOLAR2015}.

Heretofore, there have been only a few studies with detailed observations of the evolution of elemental abundances during flares \citep[e.g.,][]{narendranathCoronalElementalAbundance2020, mondalEvolutionElementalAbundances2021, namaCoronalElementalAbundances2023}.
Most previous research was restricted to comprehensive investigations of the IFIP effect at specific phases.
For example, \citet{katsudaInverseFirstIonization2020} focused only the flare peaks, whereas \citet{bakerTransientInverseFIPPlasma2019, bakerCanSubphotosphericMagnetic2020} and \citet{brooksDiagnosticCoronalElemental2018} concentrated on the decay phase.
This makes the currently proposed physical interpretations ad hoc to the individual phases.
Furthermore, although \citet{katsudaInverseFirstIonization2020} observed the depletion of low-FIP elements, such as Si and S, no analysis of Fe was undertaken. This omission is noteworthy, considering that Fe and Si have similar FIP values (Fe: 7.87~eV; Si: 8.15~eV) but their atomic masses differ significantly (Fe: 55.847~amu; Si: 28.0855~amu). It is therefore important to make further investigation and to enhance the understanding of the underlying mechanisms involved in solar flares.

Chang'E-2 (CE-2) was launched on October 1, 2010, and carries a Solar X-ray Monitor (SXM) as part of the X-ray spectrometer \citep{pengProspectiveResultsCHANG2009, banResearchInversionElemental2014a, dongCalibrationsChangE22019}.
The primary role of the SXM is to measure the spectrum of solar SXR and the solar flux.
The SXM consists of a 500~$\mathrm{\text\textmu m}$ thick Si-PIN solid-state detector with a thin 12.5~$\mathrm{\text\textmu m}$ beryllium (Be) window, which covers an energy range from 0.6 to 10.7~keV. It offers an good energy resolution of 300~eV at 5.95~keV.
The detection area is 25~cm$^2$, but a copper cap with an annular hole of central diameter, 0.5~mm, is mounted in front of the Be window so as to reduce the effective detection area by approximately 99.2\% to 0.2~mm$^2$. The purpose of this reduction is to prevent the detector count rate from saturation during solar flare events.
During a 10-second integration, the X-ray emissions are collected and recorded in 976 channels.
The detailed geometry and the specifications of the CE-2/SXM have been published previously \citep{pengProspectiveResultsCHANG2009, banResearchInversionElemental2014a, dongCalibrationsChangE22019}.

In this study, we leverage the high-resolution data obtained from the Solar X-ray Monitor on-board China's Chang'E-2 lunar orbiter \citep{pengProspectiveResultsCHANG2009, banResearchInversionElemental2014a, dongCalibrationsChangE22019} to study the elemental abundances of two solar flare events.
Our study provides a comprehensive analysis of the evolution of elemental abundances (Fe, Ca, S, Si, and Ar) as flares evolve with a high temporal cadence.
The flares under study include a Geostationary Operational Environmental Satellite (GOES) M1.3-class flare in AR 11149 (FL1) and a GOES C7.8-class flare in AR 11158 (FL2), both observed by CE-2/SXM throughout the entire flare process.
This extensive observational coverage extends beyond the limited scope of previous studies, which primarily focused on specific phases.
The interpretation derived from these expansive observations sheds light on the pivotal role of ion inertia in the FIP and IFIP effects during solar flares, in addition to the fractionation involved in Laming's ponderomotive force fractionation model.

The rest of the paper is organized as follows.
We describe the details of the observations and CE-2/SXM data analysis for the two flares in Section~\ref{sec_obs}.
In Section~\ref{sec_results}, we present the results of the evolution of elemental abundances, followed by the physical interpretation.
Finally, we summarize the paper with the discussion in Section~\ref{sec_summary}.

\section{Observations and Data Analysis}
\label{sec_obs}

\subsection{Event overview}
\label{sec_eventoveriew}

\begin{figure}[ht!]%
	\centering
	\includegraphics[width=0.7\textwidth]{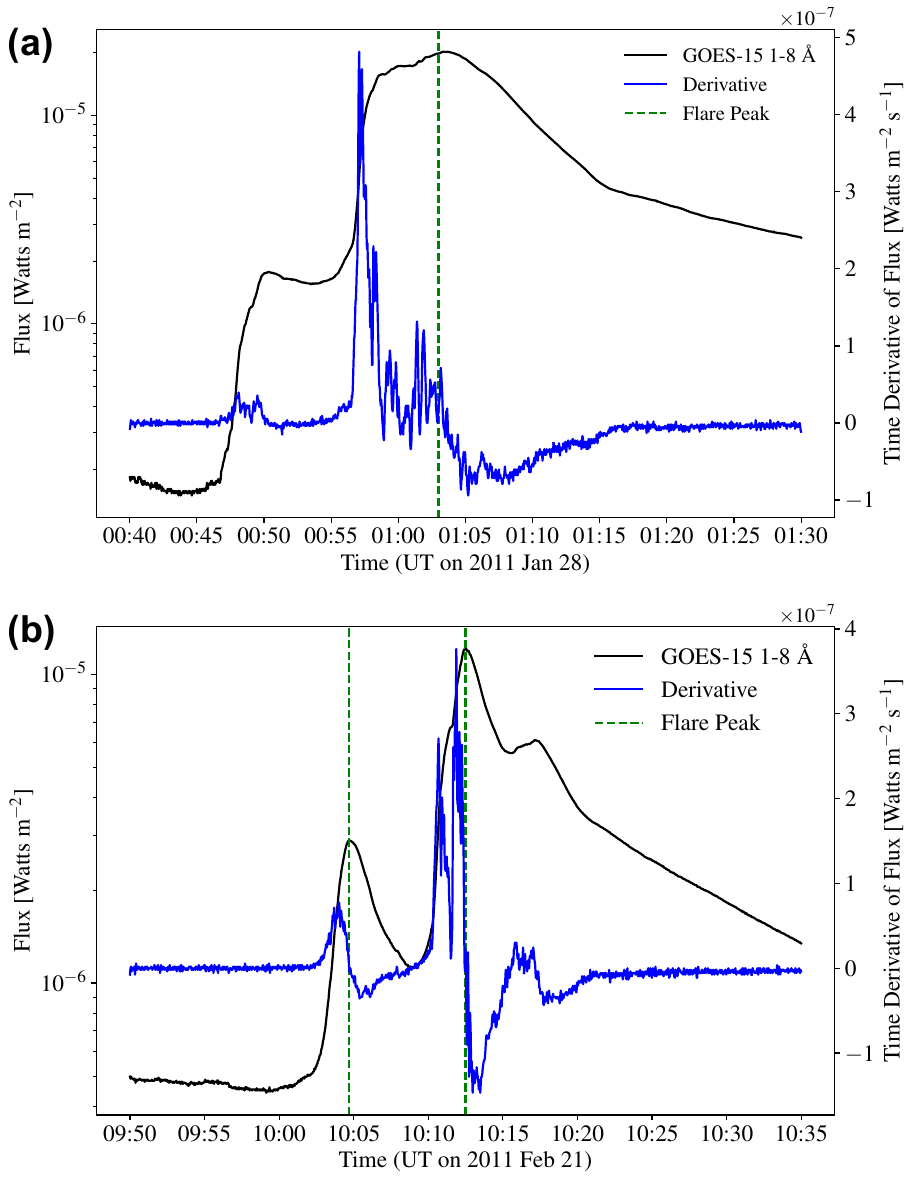}
	\caption{GOES SXR 1--8~{\AA} flux curves (in black) of flares (a) FL1 and (b) FL2, with their derivative curves (in blue). The dashed green lines represent the time of the flare peak for (a) FL1 and (b) FL2. Note that the dashed green line on the left in the panel (b) represents the flare peak for another minor flare before flare FL2.}
	\label{fig_Flux_Neupert}
\end{figure}

The first flare event (FL1) erupts on January 28, 2011, and is located at AR 11149 on the west solar limb. It starts at about 00:44~UT and peaks at about 01:03~UT (Figure~\ref{fig_Flux_Neupert}(a)).
The second flare event (FL2) erupts on February 21, 2011, and is located at AR 11158 on the west solar limb. It starts at about 10:08~UT and peaks at about 10:13~UT (Figure~\ref{fig_Flux_Neupert}(b)).
It is noted that the two flares exhibit distinct light curves, one is a long duration event (LDE) and the other is an impulsive event.
In the images obtained by the Solar Dynamics Observatory \citep[SDO;][]{pesnellSolarDynamicsObservatory2012}/Atmospheric Imaging Assembly \citep[AIA;][]{lemenAtmosphericImagingAssembly2012} in the 304~{\AA} (\ion{He}{2}, $\log T=4.7$~K) and 193~{\AA} (\ion{Fe}{12} and \ion{Fe}{24}, $\log T=6.1$--$7.3$~K) channels (Appendix Figures~\ref{fig_M1365AIA} and \ref{fig_C1663AIA}), flare FL1 is identified as an eruptive flare, whereas flare FL2 is classified as a confined flare.
It is noted that an adjacent candle-flame-shaped flare located immediately north of flare FL1 had not yet started \citep{guidoniTEMPERATUREELECTRONDENSITY2015, tsunetaStructureDynamicsMagnetic1996, forbesReconnectionFieldLine1996}.
The AIA imaging data reveal the contributions from spectral lines and continuum, enabling diagnostics of the plasma temperature and density.
The AIA 304~{\AA} channel observes emissions from \ion{He}{2}, which is formed at the typical temperature of chromosphere and transition region of the solar atmosphere ($\log T=4.7$~K).
In contrast, the AIA 193~{\AA} channel captures significant emissions from \ion{Fe}{12} ($\log T=6.1$~K) in the corona.
With the AIA imaging observations, it is noted that the origin of the flare FL1 is attributed to the filament eruption (see the animation associated with Appendix Figure~\ref{fig_M1365AIA} for flare FL1 in the 304~{\AA} and 193~{\AA}), whereas that of the flare FL2 is triggered by the magnetic reconnection originating from the interaction between emerging magnetic flux and the pre-existing magnetic field within the solar atmosphere (see the animation associated with Appendix Figure~\ref{fig_C1663AIA} for flare FL2 in the 304~{\AA} and 193~{\AA}).
On the other hand, it is noted that conducting a comparative analysis for elemental abundances between these two spectral channels remains somewhat challenging, since the radiation intensity observed in each channel correlates not only with the temperature but also with the ion density, complicating straightforward comparative assessments.

Furthermore, the magnetic field conditions for both flares FL1 and FL2 are evaluated using the data from the SDO/Helioseismic and Magnetic Imager \citep[SDO/HMI;][]{schouDesignGroundCalibration2012, scherrerHelioseismicMagneticImager2012} Space-weather HMI Active Region Patches \citep[SHARPs;][]{bobraHelioseismicMagneticImager2014}.
Prior to the eruptions of FL1 (on January 27, 2011 at 14:00~UT) and FL2 (on February 20, 2011 at 20:00~UT), which are 10 and 14 hours before each event, respectively, the estimated radial magnetic field magnitudes, $\bm{B}_r$, were $\gtrsim$1300~G for FL1 and $\gtrsim$900~G for FL2.
Both ARs contained sunspots with areas of $\geqslant$200 millionths of a solar hemisphere (MSH).
Unfortunately, the ARs of interest were situated at the solar limb, which hinders the magnetic observation of these ARs at the time of flare eruptions.

\subsection{Spectral Analysis}
\label{sec_analysis}

This study is based on the level-2B CE-2/SXM data available from the Ground and Research Application System (GRAS) of the Chang'E-2 Project\footnote{\href{https://moon.bao.ac.cn}{https://moon.bao.ac.cn}}.
A total of 2403 orbital data were collected by the CE-2/SXM instrument, covering the mission time from October 1, 2010 to May 20, 2011.
We performed a calibration process on CE-2/SXM data using the method described by \citet{dongCalibrationsChangE22019}. The calibration process includes energy-channel conversion ($\mathrm{Energy\,[eV]}=10.536\times\mathrm{Channel}-40.424$), deadtime calibration (deadtime $\tau=160$~$\mathrm{\text\textmu s}$, \citealp{dongCalibrationsChangE22019}), collimator correction, and background subtraction.
The CE-2/SXM non-solar background is obtained by summing all measurements taken during the dark side of the Moon in the periods of solar quiescence and then normalizing them by the accumulation time~\citep{dongCalibrationsChangE22019}.
This calibration procedure is essential for reducing systematic errors and enhancing the accuracy of the data. For a comprehensive calibration analysis, we refer the reader to \citet{dongCalibrationsChangE22019}.
Any data that have a ``quality state'' value other than ``00'' are excluded from the analysis.
Further, the spectra have been rebinned with a bin size of 10, giving spectral channels of 105.36~eV per bin.

The solar soft X-ray spectra typically exhibit a continuum resulting from free-free and free-bound thermal radiative processes, along with emission lines corresponding to different ionization states of the measured elements.
In the present fitting analysis, we employ the two-temperature (2-T) model {\tt vth+vth\_abun}, provided by the Object Spectral Executive~\citep[OSPEX\footnote{\href{https://hesperia.gsfc.nasa.gov/ssw/packages/spex/doc/ospex_explanation.htm}{https://hesperia.gsfc.nasa.gov/ssw/packages/spex/doc/ospex\_explanation.htm}};][]{tolbertOSPEXObjectSpectral2020}---a standard tool set for solar data analysis distributed within the SolarSoftWare~\citep[SSW;][]{freelandDataAnalysisSolarSoft1998} package.
This model is the same as the one used by \citet{dennisSOLARFLAREELEMENT2015}.
We treat the abundances of the higher temperature component ({\tt vth\_abun}) as free parameters (allowing them to vary), while the abundances of the lower temperature component ({\tt vth}) are assumed to remain constant, representing a continuum.
The selected energy range for fitting spans from 1.1 to 8.0~keV, with the lower limit set due to the need for a better detector response matrix (DRM).
Energies below 1.1~keV are primarily affected by instrumental effects, including significant attenuation due to the Be window, Si-K escape, Si-L escape, and Compton scattering, which hinder the detection of signals from the Sun.
Energies above 8.0~keV are ignored due to their larger uncertainties.
The choice of the energy range from 1.1 to 8.0~keV allows for a more accurate determination and achieves a good fit.
The CE-2/SXM count-rate spectra are fitted with a sum of two isothermal model spectra of solar photon intensity versus energy, generated using the CHIANTI atomic database version 9.0.1 \citep{dereCHIANTIAtomicDatabase1997, dereCHIANTIAtomicDatabase2019}.
These isothermal model spectra are calculated based on functional forms that incorporate free parameters, such as temperature ($T$), emission measure ($\mathrm{EM}=N_e^2\,V$, where $N_e$ is the electron density and $V$ is the emitting volume), elemental abundances (e.g., Fe, Ca, S, Si, and Ar), and ionization fractions \citep{mazzottaIonizationBalanceOptically1998}.
In our analysis, the coronal and photospheric abundances on an absolute scale with respect to hydrogen (H) are chosen according to \citet{feldmanElementalAbundancesUpper1992} and \citet{asplundChemicalCompositionSun2009}, respectively.
The goodness-of-fit is characterized quantitatively by the reduced chi-squared statistic, which is calculated as
\begin{equation}
	\chi_\mathrm{red}^2 = \frac{1}{\nu}\sum_i\left[\frac{(C_{i,\mathrm{obs}}-C_{i,\mathrm{model}})^2}{\sigma_{i,\mathrm{model}}^2}\right],
\end{equation}
where $C_{i,\mathrm{obs}}$ and $C_{i,\mathrm{model}}$ are the measured and model count rates, respectively, in the $i^\mathrm{th}$ energy bin, $\sigma_{i,\mathrm{model}}$ indicates the standard deviation of the model count rate assuming Poisson counting statistics in the $\chi^2$ calculation, and $\nu$ represents the number of degrees of freedom.

\section{Results}
\label{sec_results}

\subsection{Evolution of elemental abundances and physical interpretation}
\label{sec_evolution}

\begin{figure}[ht!]%
	\centering
	\includegraphics[width=0.55\textwidth]{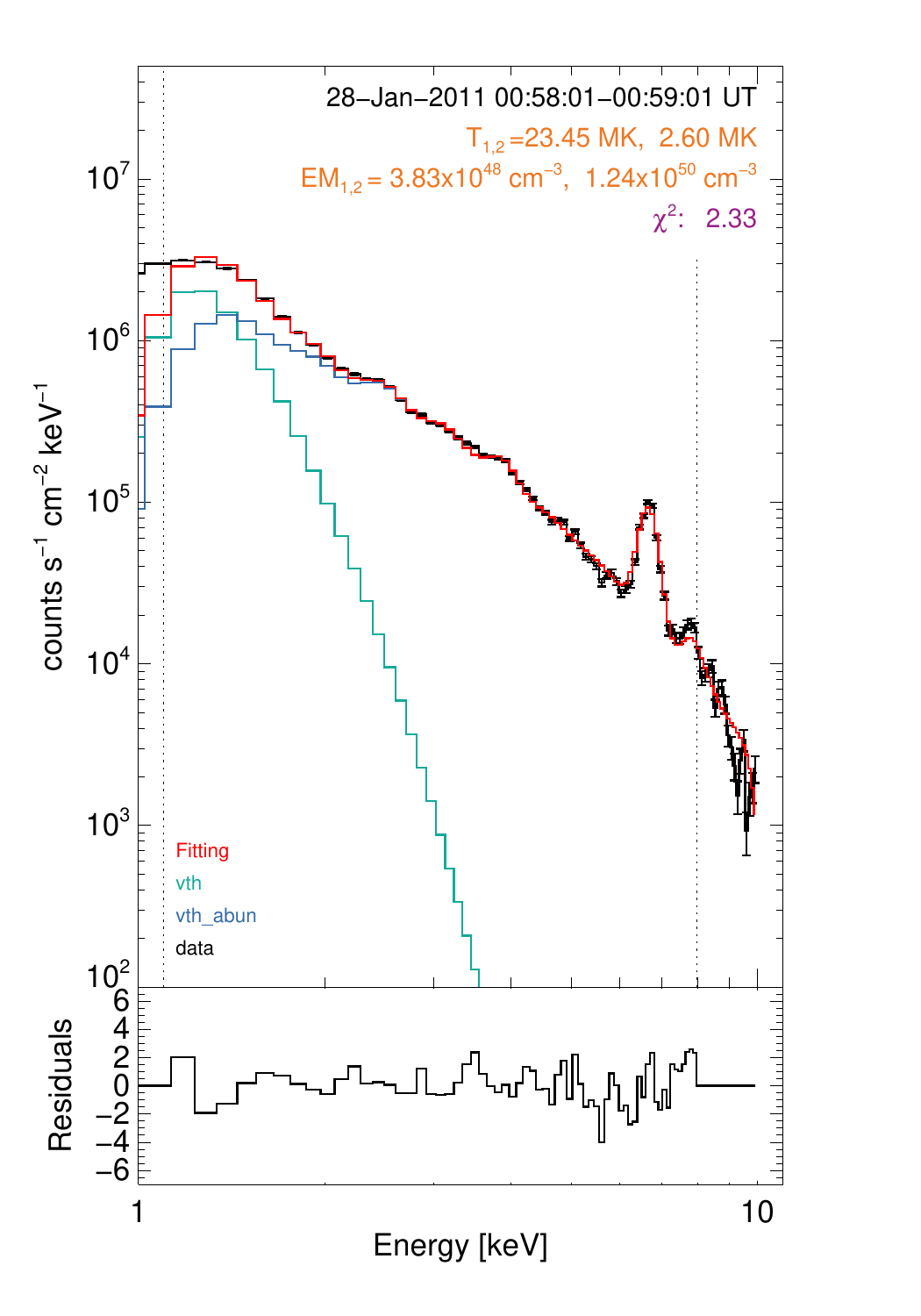}
	\caption{A representative 2-T model fit (red curve) of FL1 (M1.3) for the 00:58:01--00:59:01~UT period on January 28, 2011, with normalized residuals, which are defined as the ratio of the differences between the measured and best-fit fluxes to the 1$\sigma$ statistical uncertainties.
	The CE-2/SXM count flux background-subtracted spectrum is represented by black histograms.
	The red, emerald-green, and blue colored curves correspond to the total fit 2-T model, the lower temperature component ({\tt vth}), and the higher temperature component ({\tt vth\_abun}), respectively.
	}
	\label{fig_FitPlot_1365_55}
\end{figure}

For both flares, we obtained the best-fit spectral parameters of the emission measure, temperature, and abundance with $\pm1\sigma$ uncertainties by fitting the solar soft X-ray spectra of CE-2/SXM using a two-temperature (2-T) fit function through the OSPEX software package \citep{tolbertOSPEXObjectSpectral2020} (see Section~\ref{sec_analysis}).
Figure~\ref{fig_FitPlot_1365_55} shows the representative CE-2/SXM count-rate spectrum of FL1 for the 00:58:01--00:59:01~UT period on January 28, 2011, which approximately corresponds to the temperature peak, in units of counts~s$^{-1}$~cm$^{-2}$~keV$^{-1}$, together with the result of a 2-T model spectral fit in the upper panel.
The normalized residuals, defined as the ratio of the differences between the measured and best-fit fluxes to the 1$\sigma$ statistical uncertainties, are shown in the lower panel.
The spectral fit reveals that the higher temperature component has a temperature of $T=23.45$~MK ($=2.02$~keV) and an emission measure of $\mathrm{EM}=3.83 \times 10^{48}$~cm$^{-3}$, representing the hot flare plasmas; the lower temperature component has a temperature of $T=2.60$~MK ($=0.22$~keV) and an emission measure of $\mathrm{EM}=1.24 \times 10^{50}$~cm$^{-3}$, representing the plasmas of the ``typical'' quiet corona; and the goodness-of-fit is characterized by $\chi_\mathrm{red}^2=2.33$.
The background-subtracted spectrum is dominated by the continuum emission.
The presence of Fe and Fe/Ni line complexes is clearly discernible at $\sim$6.7~keV and $\sim$8~keV, respectively.
Moreover, the contributions from various elements exhibit peak intensities at the following energies: Fe at 1.5~keV, 6.7~keV, and 8~keV; Ca at 3.9~keV; S at 2.5~keV; Si at 2.0~keV; Ar at 3.2~keV; and Mg and other elements at around 1.6~keV (cf.~Figure~3 of \citealp{dennisSOLARFLAREELEMENT2015}).

\begin{figure}[ht!]%
	\centering
	\includegraphics[width=0.55\textwidth]{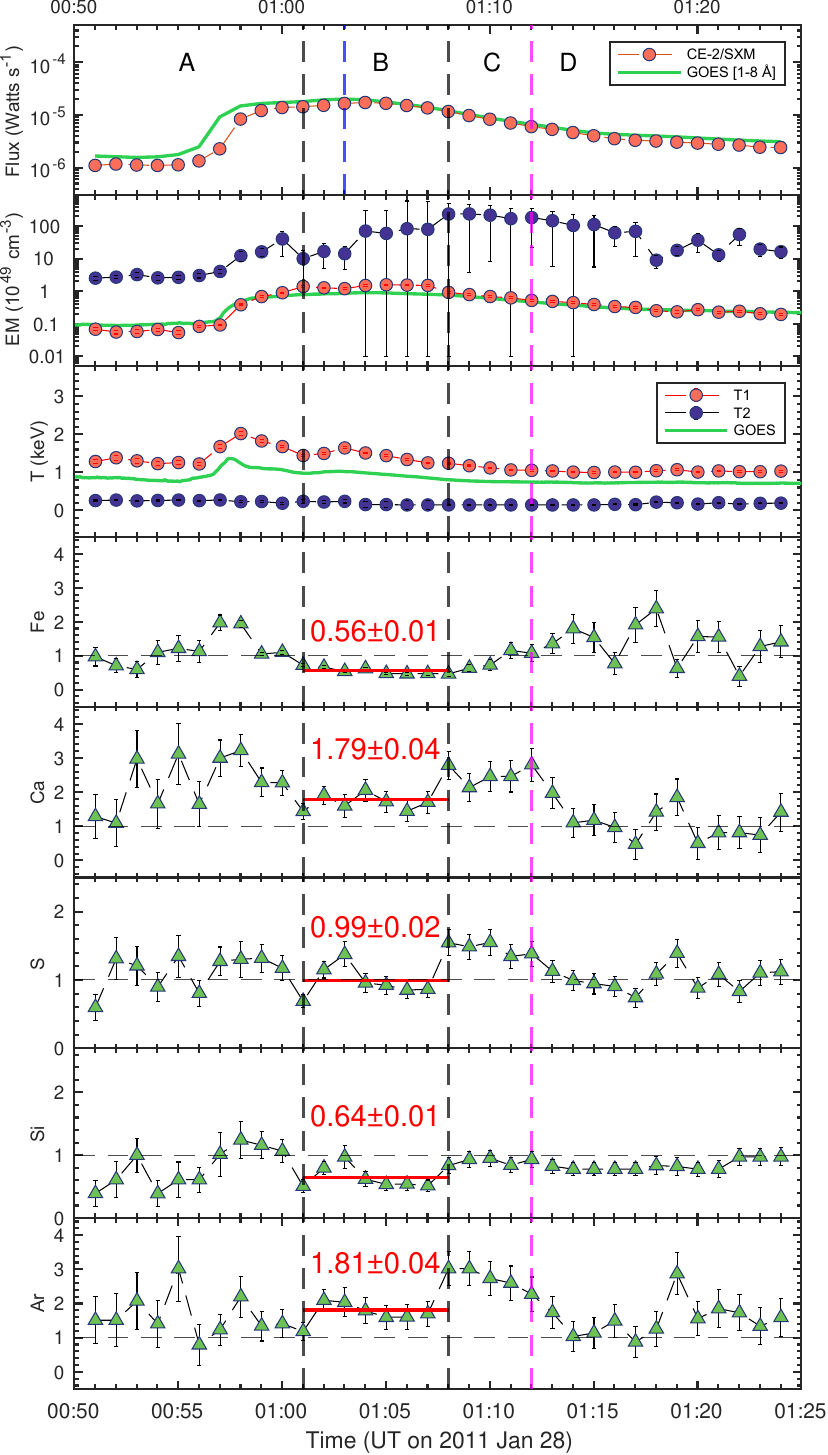}
	\caption{Soft X-ray 1--8~{\AA} fluxes, estimated emission measures, temperatures, and abundances with $\pm1\sigma$ uncertainties obtained from CE-2/SXM spectra using the two-temperature (2-T) fit function for FL1.
	Panels 1, 2, and 3 show the evolution of soft X-ray 1--8~{\AA} fluxes (in Watts~s$^{-1}$), emission measures, and temperatures (in keV), respectively.
	In panel 1, the red and green curves represent the SXR 1--8~{\AA} fluxes observed by CE-2/SXM and GOES 15 XRS-B [1--8~{\AA}], respectively, with the dashed blue vertical line indicating the time when flux reaches its peak.
	In panels 2 and 3, the red and blue colored curves represent the best-fit parameters obtained from the fitted spectra of CE-2/SXM of $T_1$ (hot component) and $T_2$ (cool component). The green curve represents the measurements from GOES 15.
	Panels 4 to 8 show the evolution of FIP biases for Fe, Ca, S, Si, and Ar, respectively.
	The FIP bias is defined as the ratio of the measured elemental abundance with respect to the photospheric abundance, where the coronal and photospheric abundances are on an absolute scale (with respect to H) and taken from \citet{feldmanElementalAbundancesUpper1992} and \citet{asplundChemicalCompositionSun2009}, respectively.
	The y-error bars represent 1$\sigma$ uncertainties for each of the parameters, and the x-error bars represent the duration over which a spectrum is integrated.
	The left-dashed black vertical line indicates the time when EM1 reaches its peak \citep[cf.][]{mondalEvolutionElementalAbundances2021}, while the right-dashed vertical black line indicates the time when $T_1$ decreases smoothly.
	The dashed magenta vertical line denotes the beginning when occurs the subsequent depletion.
	The horizontal red lines show the weighted average FIP biases over the indicated time intervals, and the values are marked.
	The letters A, B, C, and D correspond to the impulsive phase (A), the peak phase (B), and the decay phase (C and D), respectively.
	}
	\label{fig_M1365_FIP}
\end{figure}

\begin{figure}[ht!]%
	\centering
	\includegraphics[width=0.6\textwidth]{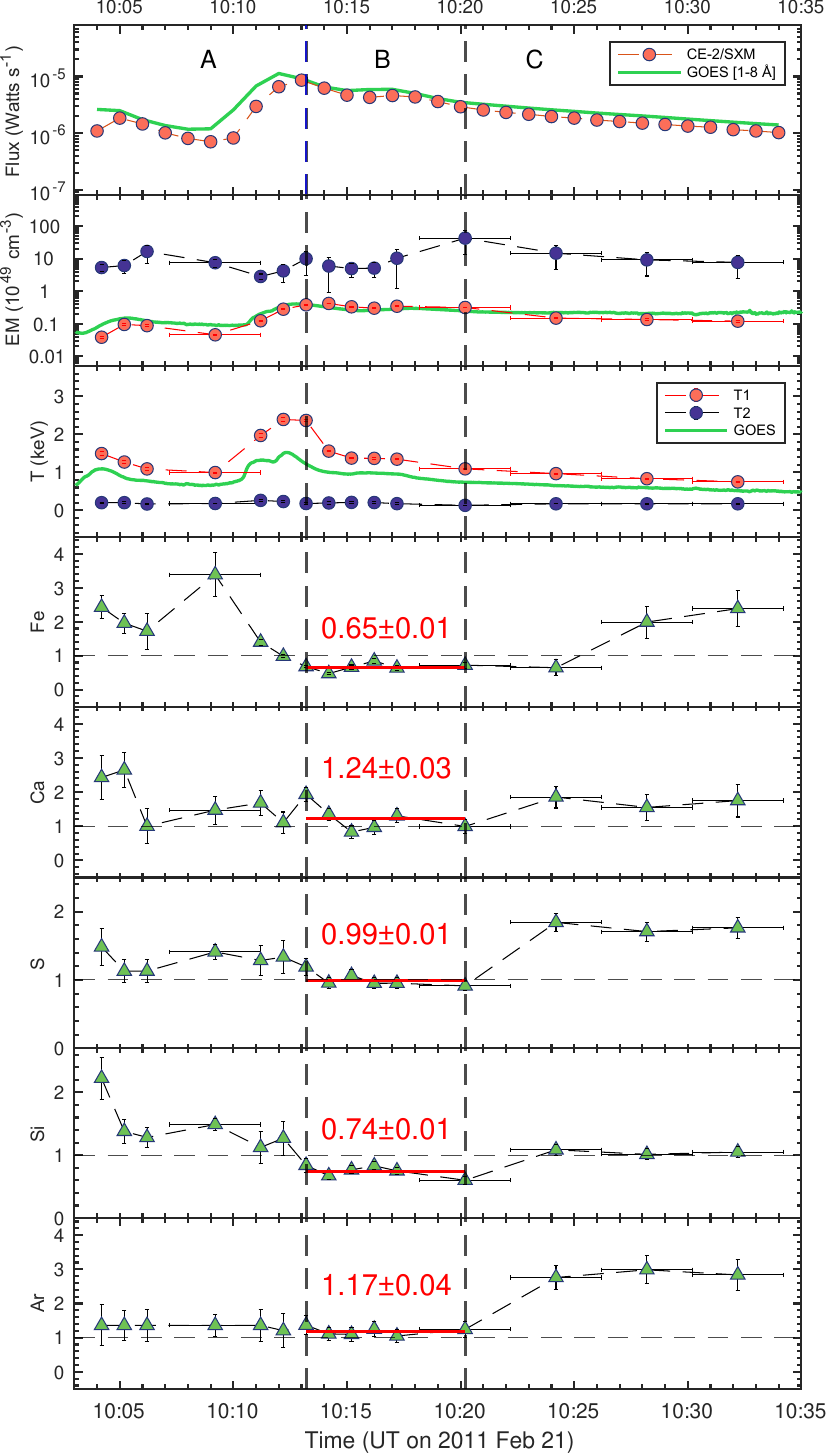}
	\caption{Soft X-ray 1--8~{\AA} fluxes, estimated emission measures, temperatures, and abundances with $\pm1\sigma$ uncertainties obtained from CE-2/SXM spectra using the two-temperature (2-T) fit function for FL2. The indications are the same as those of Figure~\ref{fig_M1365_FIP}.}
	\label{fig_C1663_FIP}
\end{figure}

The soft X-ray 1--8~{\AA} fluxes, emission measures, and temperatures of flares FL1 and FL2 are depicted in the top three panels of Figures~\ref{fig_M1365_FIP} and \ref{fig_C1663_FIP}, respectively.
The remaining panels illustrate the evolution of the FIP bias of elemental abundances, defined as the ratio of the measured elemental abundance in the flare to the photospheric abundance.
It is seen that the two flares represent two distinct scenarios in terms of the elemental (Fe, Ca, S, Si, and Ar) evolution: (1) most typical cases (corresponding to FL2, Figure~\ref{fig_C1663_FIP}) and (2) rare cases (corresponding to FL1, Figure~\ref{fig_M1365_FIP}).
The entire flare process is divided into different phases: the impulsive phase (marked as A), the peak phase (marked as B), and the decay phase (marked as C and D).
These phases will be extensively explored in the subsequent subsections~\ref{sec_initial}--\ref{sec_subdecay}.

\subsubsection{Pre-impulsive \& Impulsive phases}
\label{sec_initial}

In the quiet Sun before the flare onset, the elemental abundances in the corona are usually enhanced by a factor of 3--4 relative to those in the photosphere.
This enhancement is prominent for the low-FIP (below $\sim$10~eV) elements like Mg, Fe, and Si, which can be photoionized by \ion{H}{1} Ly$\alpha$ and are primarily ionized in the chromosphere. In contrast, high-FIP elements like Ar, O, and S tend to remain relatively unaffected.
The phenomenon can be clearly explained by the sophisticated fractionation mechanism (see Section~\ref{sec_intro}) which involves ion-neutral separation in the solar chromosphere \citep{lamingUnifiedPictureFirst2004, lamingTHERMALCONDUCTIVITYELEMENT2009, lamingNONWKBMODELSFIRST2009, lamingNONWKBMODELSFIRST2012, lamingFIPInverseFIP2015, lamingFirstIonizationPotential2017, lamingElementAbundancesNew2019}.
The FIP effect arises due to the upward ponderomotive force generated by upward propagating Alfvén waves, originating either from the photospheric and/or from the corona and then reflected.
This force acts on the low-FIP elements in the chromosphere, causing fractionation. Consequently, the coronal plasma becomes enriched with low-FIP elements.

In Figures~\ref{fig_Flux_Neupert}(a--b) we plot the time-derivative of the soft X-ray flux of flares FL1 and FL2, as indicated by the blue curves.
According to the empirical Neupert effect, they should be concurrent with the hard X-ray light curves \citep{neupertComparisonSolarXRay1968, veronigPhysicsNeupertEffect2005}, and the chromospheric evaporation plays a direct role for such a relationship \citep{ningINVESTIGATIONCHROMOSPHERICEVAPORATION2010}.
According to the CSHKP model (see Section~\ref{sec_intro}), as the flare starts, thermal flux and non-thermal electrons from the reconnection site are channeled downward to the chromosphere, fostering the formation of a high-pressure, high-temperature region.
This temperature surge induces the heated plasma to undergo upward evaporation, driven by the pressure gradient.
This evaporation drives a significant mass transfer from the chromosphere into the corona.
Examining Figures~\ref{fig_M1365_FIP} and \ref{fig_C1663_FIP} during the period marked as A, the flares start with both temperature and emission measures gradually increasing.
Their peaks occur during the impulsive phase, with the temperature typically peaking before the emission measure. This temporal sequence indicates that plasma heating commences prior to the expansion of the emitting material.
As energy is transferred to the flare loop footpoints, these regions radiate hard X-ray emissions through non-thermal bremsstrahlung, coinciding with the near-simultaneous evaporation of plasma enriched low-FIP elements above the chromosphere during the flare onset.
It should be noted that during the pre-flare phase of the flare FL2, there was a small flare eruption, so the elemental abundances decreased to the photospheric values.

\subsubsection{Peak phase}
\label{sec_peak}

At the next stage, designated as the peak phase (marked as B) from 01:01~UT to 01:08~UT in Figure~\ref{fig_M1365_FIP} and from 10:13~UT to $\sim$10:22~UT in Figure~\ref{fig_C1663_FIP}, a clear depletion of abundances is observed for Ca, Fe, and Si, with weighted average FIP biases of 1.79, 0.56, 0.64, respectively, for FL1, and 1.24, 0.65, 0.74, respectively, for FL2.
This suggests that Fe and Si exhibit the IFIP effect during the peak phase, with Fe and Si abundances falling to their photospheric values in both flares.

This phase coincides with the completion of the evaporation from the deep chromosphere.
This is notably evident as the soft X-ray flux reaches its peak during this phase, marked as flare peak by dashed blue vertical line in Figures~\ref{fig_M1365_FIP} and \ref{fig_C1663_FIP}.
In this phase, plasma originating from the deeper chromosphere begins to evaporate.
As a consequence, the rapid upward speed hinders fractionation during plasma evaporation \citep{lamingNONWKBMODELSFIRST2009}.
The plasma from the lower chromosphere, with less ionized low-FIP elements \citep{warrenMEASUREMENTSABSOLUTEABUNDANCES2014} (hereafter referred to as IFIP effect plasma), leads to a rapid decrease in the abundances of low-FIP elements.
They tend to approach values close to their photospheric levels during this phase.

\citet{katsudaInverseFirstIonization2020} reported that around flare peaks, the depletions of Si and Ca show average values of $\sim$0.7 and $\sim$2.0 times their photospheric abundances, respectively, which indicates that our results are in good agreement with their observations of large flares \citep{katsudaInverseFirstIonization2020} and reaffirms the existence of the IFIP effect for Si.
Their analysis focused on giant X-class flares, which may account for the slight difference in their value for Ca compared to our results since our events are weaker.
\citet{lamingFIPInverseFIPEffects2021} presented a more detailed picture of FIP fractionation. According to his model, near the top of the chromosphere, the FIP effect acts on different ions in different degrees, depending on their ionization fractions. This difference in fractional degrees results in an overall FIP effect for Ca and an IFIP effect for Si and S~\citep{katsudaInverseFirstIonization2020, lamingFIPInverseFIPEffects2021}, consistent with our observational results.

Whereas \citet{katsudaInverseFirstIonization2020} observed that the intermediate-FIP element S (FIP: 10.36~eV) exhibited the IFIP effect, with an elemental abundance of $\sim$0.3 times its photospheric abundance in giant X-class flares, our observations revealed that the weighted average FIP bias of S (around 1) remained nearly constant from the impulsive phase to the peak phase, albeit showing relatively slight depletion during the peak phase.
Note that sulfur, as an intermediate-FIP element, has a unique hybrid nature, behaving as both high-FIP and low-FIP elements.
Our observations suggest that sulfur behaves more like a low-FIP element in modest (upper C--M class) flares; conversely, in weak (A--C class) flares, sulfur behaves like a high-FIP element in the manner described by \citet{narendranathElementalAbundancesSolar2014}.
It is plausible that sulfur behaves like a high-FIP element in weak (A--C class) flares while exhibiting characteristics of a low-FIP element in strong (M--X class) flares.
However, more observational evidence is needed to confirm this conclusively.
Furthermore, as indicated by \citet{lamingFIPInverseFIPEffects2021}, the ionization fraction of sulfur is approximately 94\% in the upper chromosphere, decreasing to around 85\%--90\% near the equipartition layer.
The modeling of the IFIP effect for sulfur using this profile \citep{lamingFIPInverseFIPEffects2021} agrees with the work by \citet{katsudaInverseFirstIonization2020}, given the magnetic field expansion from the photosphere to the corona (expressed as $B_\mathrm{cor}/B_\mathrm{photo}$) of 0.7 with a fast-mode wave amplitude ($\nu_\mathrm{fm}$) at the equipartition layer of 7~km~s$^{-1}$.
This model suggests that sulfur is less depleted compared to various low-FIP elements (Ca, Fe, and Si) above the equipartition layer but is most depleted at the top of the chromosphere, where it is not completely fractionated (around 94\%).
Our observational results agree with the calculations of \citet{lamingFIPInverseFIPEffects2021} for $B_\mathrm{cor}/B_\mathrm{photo}=0.7$, with relatively fluctuated values around this level in our study and a fast-mode wave amplitude $\nu_\mathrm{fm}$ at the equipartition layer higher than 6~km~s$^{-1}$. However, our observations of sulfur may rather favor a sulfur ionization fraction above 99\% in the upper chromosphere, particularly at the top, similar to what described in \citet{lamingElementAbundancesNew2019}.
Of particular interest is integrating our interpretation with the ionization fraction profile for S, which provides clues about the depth from which the plasma, brought by evaporation to the corona or flare loop, originates.
The IFIP effect plasma appears more prominently, driven by the higher energy released during strong flares, which penetrates deeper into the chromosphere, compared to weaker flares.
In contrast, weak flares are unable to drive the deep IFIP effect plasma in the lower chromosphere upward, affecting only the fractionated FIP effect plasma in the upper chromosphere.
This understanding is key to correlating flare intensity, sulfur's behavior in FIP/IFIP effects, and the plasma's source height.

Since the IFIP effect for Fe in our work is discovered for the first time, it provides us with new insight into the underlying physical mechanisms and should be of great interest to the entire solar physics community.
It should be noted that the best-fit parameter for Fe is determined with high confidence, as the Fe peak is the most clearly defined among all observed peaks.
Hence, our observations distinctly demonstrate the depletion of low-FIP elements, particularly exhibiting the IFIP effect in the main phase of flares.
This depletion stands as a crucial characteristic of the plasma dynamics during solar flares, offering valuable insights into the processes within the chromosphere and corona during these events.

\subsubsection{Decay phase}
\label{sec_decay}

The elemental abundances swiftly recover and return to their pre-flare coronal levels during the decay phase (marked as C from 01:08~UT to 01:12~UT in Figure~\ref{fig_M1365_FIP} and from $\sim$10:22~UT to 10:34~UT in Figure~\ref{fig_C1663_FIP}). 

As the upflow speed significantly decreases with declining temperature, plasma begins to revert to fractionation.
The fractionated FIP effect plasma in the upper chromosphere refills the flare loop and replaces the IFIP effect plasma, resulting in the observed abundance enhancements.
This is consistent with previous observations conducted by \citet{doschekPhotosphericCoronalAbundances2018}, \citet{bakerPlasmaCompositionSigmoidal2013, bakerTransientInverseFIPPlasma2019, bakerCanSubphotosphericMagnetic2020}, \citet{mondalEvolutionElementalAbundances2021}, and \citet{toEvolutionPlasmaComposition2021}. These studies focused on flares associated with loop-like structures, a feature also exhibited in FL2.
Similarly, the plasma replacement is the cause of the recovery to the pre-flare coronal abundances for flare FL1 as well.

As time progresses, the dominance of evaporation gradually diminishes, giving way to a more significant role played by thermal conduction. This transition may also involve the interplay of radiative cooling and gradual magnetic reconnection \citep{bothaThermalConductionEffects2011, tamMHDSimulationsCoronal2014, reale3DMHDMODELING2016}.
The generation of coronal Alfvén waves resulting from magnetic reconnection \citep[e.g.,][]{fletcherImpulsivePhaseFlare2008} in the solar corona could serve as an additional driver contributing to the increases in elemental abundances, where the condition required is that the Alfvén speed should be faster than the thermal electron speed, as suggested by \citet{toEvolutionPlasmaComposition2021} and \citet{mondalEvolutionElementalAbundances2021} (the second possibility).
The coronal Alfvén waves, produced by magnetic reconnection and initiated at the flare onset, propagate downward from the flaring site to the lower solar atmosphere. Upon reaching the lower chromosphere, these waves, being slower than the arrival of heat conduction flux carried by the suprathermal electrons (which are near-relativistic in speed; see Section~\ref{sec_intro}), are reflected, generating the upward ponderomotive forces that can contribute to the observed increase in the abundances of the low-FIP elements.
It is noted that the propagating of coronal Alfvén waves into the lower solar atmosphere has not been directly observed yet, and further observation is required. Such observations would aid in the quantitative modeling of the role played by these waves in altering the elemental abundances over time during flares.

\begin{figure}[ht!]%
	\centering
	\includegraphics[width=0.8\textwidth]{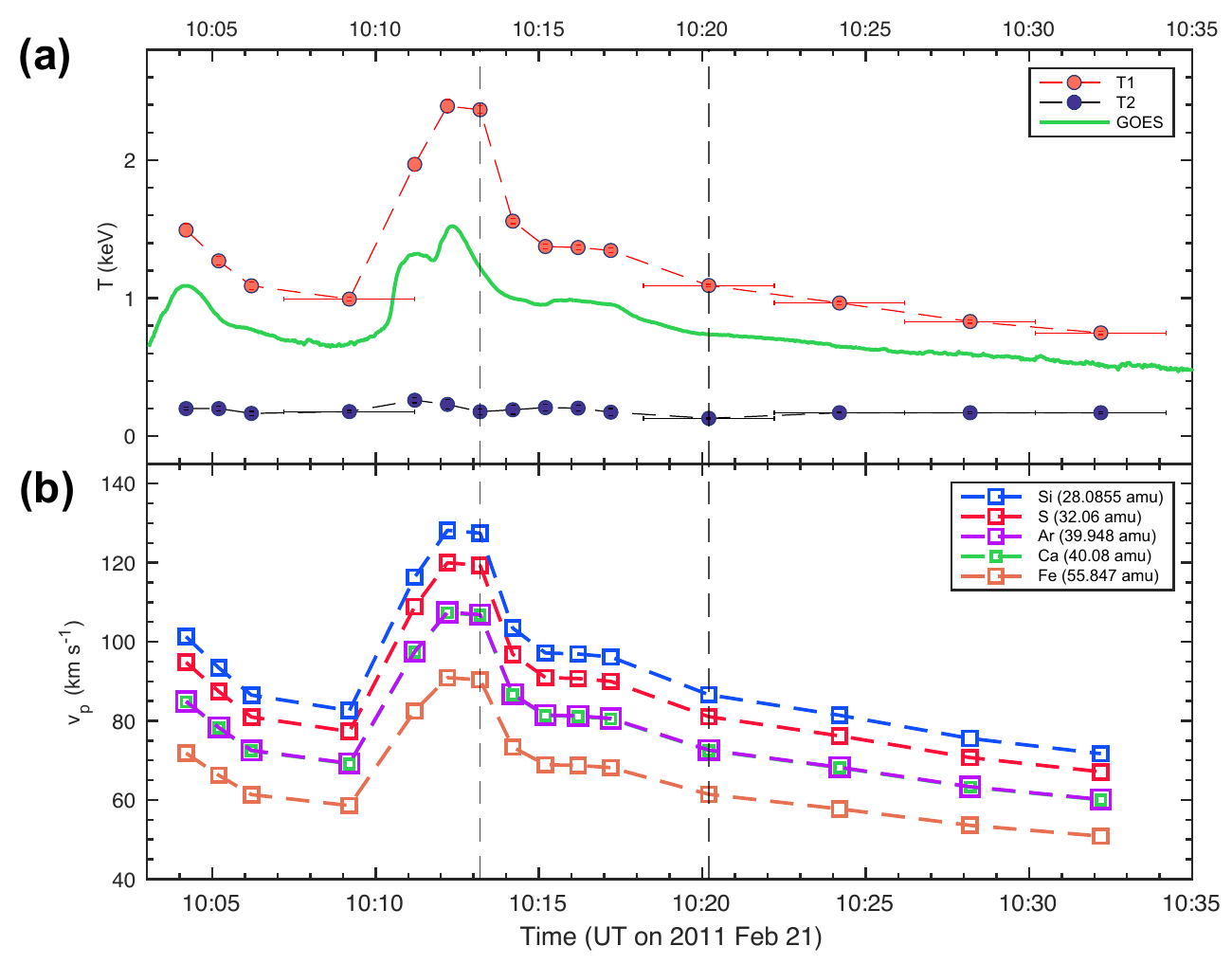}
	\caption{The theoretical upflow speeds for flare FL2 (representing the most typical cases).
	The panel (a) corresponds to the third panel in Figure~\ref{fig_C1663_FIP}.
	In panel (b), the orange, green, red, blue, and purple curves represent the theoretical upflow speeds (the most probable molecular speeds) of Fe, Ca, S, Si, and Ar, respectively.
	}
	\label{fig_C1663_velocity}
\end{figure}

Interestingly, during the decay phase of flare FL1 (as depicted from 01:08~UT to 01:12~UT in Figure~\ref{fig_M1365_FIP}), we observe a distinct behavior in the abundance of Fe, which shows a slower increase compared to all other elements.
It is pertinent to note that the CE-2/SXM spectra primarily capture emissions from the collisionally ionized diffuse gas.
Under the assumptions of ionization equilibrium and the proportional relationship between the upflow velocity of each element and the most probable molecular velocity $v_p$, formulated as $v_p=\sqrt{2RT/M}$, where $R=8.31$~J~mol$^{-1}$~K$^{-1}$ denotes the gas constant, $M$ denotes the molar mass of the gas, and $T$ represents the absolute temperature, we explore the implications.
In Figure~\ref{fig_C1663_velocity}, the top panel displays the temperature derived from the spectral fitting of the 2-T model (referred to as $T_1$ in Figure~\ref{fig_C1663_FIP} for flare FL2, representing the most typical cases). Meanwhile, the bottom panel illustrates the most probable molecular velocity $v_p$, calculated based on the derived temperature ($T_1$), serving as a representation for the theoretical upflow velocity of each element.
The discrepancy in atomic masses among elements strongly implies the crucial role of inertia at play in the evaporation process during flares.
This noteworthy correlation further bolsters our interpretation, offering a plausible explanation for both the FIP and IFIP effects.
Moreover, we notice that this behavior is more apparent in flare FL1 than in flare FL2 (illustrated for a corresponding period from $\sim$10:22~UT to 10:34~UT in Figure~\ref{fig_C1663_FIP}).
By integrating AIA imaging observations (Appendix Figures~\ref{fig_M1365AIA} and \ref{fig_C1663AIA}; see Section~\ref{sec_eventoveriew}), we can postulate that this discernible discrepancy in the slower response behavior may correlate with the differing structural characteristics between the two flares FL1 (eruptive flare) and FL2 (confined flare).
This suggests a relatively reduced plasma replacement in flare FL1 compared to what was observed in flare FL2.
This distinction in the plasma dynamics between the two flares offers an opportunity to investigate the transport and replacement of different compositions between various atmospheric layers during the decay phase.

We further interpret the inertia effect by analyzing the change in FIP biases between the peak phase and the decay phase ($\Delta F=\mathrm{FIP}_\mathrm{decay}-\mathrm{FIP}_\mathrm{peak}$).
The weighted average FIP biases for FL1 (from 01:09 to 01:11~UT, representing the middle interval of the decay phase) are as follows: 0.78 for Fe, 2.34 for Ca, 1.46 for S, 0.91 for Si, and 2.78 for Ar.
Comparing Fe (FIP: 7.87~eV) and Si (FIP: 8.15~eV), which share similar FIP values but differ in atomic masses, we find that the ratio of their changes ($\Delta F_\mathrm{Fe}/\Delta F_\mathrm{Si}$) is 0.8. This certainly indicates a slower response from Fe due to its stronger inertia.
Additionally, for elements Ca (FIP: 6.11~eV; 40.08~amu) and Ar (FIP: 15.76~eV; 39.948~amu), despite having similar atomic masses but differing FIP values, the ratio of their changes ($\Delta F_\mathrm{Ca}/\Delta F_\mathrm{Ar}$) equals 0.6.
This observation appears to contradict the argument regarding the inertia effect, but it does not, as we explain below.
As seen in Figure~\ref{fig_M1365_FIP}, the abundance of the high-FIP element Ar remains nearly constant from the impulsive phase to the peak phase.
Besides, as we discussed the ionization fraction for sulfur in Section~\ref{sec_peak}, it is relevant to extend the discussion to low-FIP element calcium (Ca) and high-FIP element argon (Ar) in the flare dynamics.
The ionization fraction of Ar increases steeply with height at the top of the chromosphere ($\gtrsim$1900~km above the photosphere; cf. Figure~3c of the paper), while being almost unaffected above the equipartition layer, with a value close to 0\%.
In contrast, either in the lower or upper chromosphere, the ionization fraction of Ca remains almost $\gtrsim$99\% throughout, while at the top of the chromosphere, it becomes completely fractionated.
We pointed out that during the peak phase (see Section~\ref{sec_peak}), the IFIP effect plasma brought up by evaporation originates from the deep chromosphere, while during the decay phase, the fractionated FIP effect plasma in the upper chromosphere begins to replace the IFIP effect plasma already contained in the corona or flare loop.
Therefore, the comparable difference in their fractional degrees with varying height is the possible cause of the decline in the overall change ratio for Ca and Ar ($\Delta F_\mathrm{Ca}/\Delta F_\mathrm{Ar}$), which is smaller than 1.

In summary, the observed increase in the elemental abundances during the decay phase can be attributed to the plasma replacement as the evaporation becomes more gradual (with the dominant process gradually altering, as mentioned above).
Besides, the observations strongly imply that inertia plays an important role in the evaporation process during flares.
In addition, the presence of coronal Alfvén waves could contribute, although their specific impact might vary depending on the flare structure.
Notably, the differing recovery or response in the FIP biases likely stems from the diverse evaporation/mass upflow speeds exhibited by various elements.
Moreover, the overall manifestation of FIP or IFIP effects is intricately tied to the distinct ionization fraction of these elements, as detailed in the work by \citet{lamingFIPInverseFIPEffects2021}.

\subsubsection{Subsequent depletion during decay phase}
\label{sec_subdecay}

A unique observation in FL1, spanning from 01:12~UT to 01:18~UT (marked as D) in Figure~\ref{fig_M1365_FIP} and beginning at the time indicated by the dashed magenta vertical line, showcases a rare occurrence that is a subsequent depletion of elemental abundances (with Ca, S, Ar, with Ca being the most prominent), descending to their photospheric or subphotospheric values after the rapid recovery of elemental abundances during the decay phase of FL1.
This observation vividly illustrates the pronounced IFIP effect, particularly evident in calcium (Ca).
This subsequent depletion emerges around 10 minutes after the flare peak (marked by dashed blue vertical line in Figure~\ref{fig_M1365_FIP}), which is consistent with the observations made by \citet{bakerTransientInverseFIPPlasma2019, bakerCanSubphotosphericMagnetic2020}.
Both our observations and those by \citet{bakerTransientInverseFIPPlasma2019, bakerCanSubphotosphericMagnetic2020} suggest a coherent timescale for the emergence of the IFIP effect, occurring several to tens of minutes later than the flare peak.

Previous studies \citep{doschekAnomalousRelativeAr2015, doschekMYSTERIOUSCASESOLAR2016, doschekSunspotsStarspotsElemental2017, bakerTransientInverseFIPPlasma2019, bakerCanSubphotosphericMagnetic2020} have extensively delved into the IFIP effect observed during solar flares occurring within sunspot conditions. Their focus on examining the Ar/Ca intensity ratio, whose enhancement indicates the presence of the IFIP effect, has been pivotal.
\citet{bakerTransientInverseFIPPlasma2019, bakerCanSubphotosphericMagnetic2020} revealed the appearance of plasmas exhibiting a photospheric-like composition, which strongly indicates the presence of the IFIP effect.
Our observation during the decay phase of flare FL1 (marked as D in Figure~\ref{fig_M1365_FIP}) aligns with this, clearly revealing plasmas with a photospheric-like composition, akin to Ca, S, Si, and Ar near their FIP biases of 1.
In addition, \citet{bakerTransientInverseFIPPlasma2019} and \citet{brooksDiagnosticCoronalElemental2018} suggested that this depletion may persist for as long as $\sim$40 minutes.
That is to say, the depletion in our event may potentially extend even further than 10~minutes.
However, limited data availability and lower spectral count rates prevent us from conducting longer observations of this depletion.
As a result, our current observations lack extensive data on the evolution of elemental abundances during the persistence of the IFIP effect plasmas.
Detailed and prolonged observations of these reconnection phenomena are crucial for gaining a deeper understanding of this process.

We suggest that this subsequent depletion is due to the upward transfer of the bulk plasmas from the lower chromosphere.
It occurs after the FIP effect plasma is exhausted due to the material evaporating in the upper chromosphere (as discussed in Section~\ref{sec_decay} Decay phase).
Subsequently, the plasma originating from the lower chromosphere with normal abundances begins to ascend.
The exhaustion of the FIP effect plasma may be correlated with both the flare type and whether the plasma is confined by open magnetic fields or within closed magnetic loops \citep[e.g.,][]{lamingUnifiedPictureFirst2004}.
In terms of the flare types, flare FL1 is characterized as a long duration event, whereas flare FL2 is an impulsive event (see Section~\ref{sec_eventoveriew}).
The main distinction between them lies in their temperature profiles, which is highly correlated with evaporation, as discussed in Sections~\ref{sec_peak}--\ref{sec_decay}.
During the impulsive phase, flare FL2 exhibits higher temperatures compared to flare FL1, albeit over a shorter duration ($T_1$ indicated by the red curve in Figures~\ref{fig_M1365_FIP} and \ref{fig_C1663_FIP}; see Section~\ref{sec_peak} Peak phase). Moreover, during the decay phase, flare FL2 experiences a significantly faster decrease in temperature compared to flare FL1 (see Section~\ref{sec_decay} Decay phase).
This difference in temperature profiles suggests that the long-lasting high temperature affects the evolution of the plasma composition, leading to the exhaustion of the FIP effect plasma and the upward transfer of plasmas from the lower chromosphere.
This process contributes to the occurrence of the subsequent depletion of elemental abundances.
The flare morphological configuration also has some effects on this depletion and will be discussed later in this subsection.
In addition, this observed depletion may be associated with subchromospheric or subphotospheric reconnection, as proposed by \citet{bakerTransientInverseFIPPlasma2019, bakerCanSubphotosphericMagnetic2020}, especially given the close proximity of sunspots to flare FL1.
Highly complex magnetic fields, especially in the umbrae of coalescing sunspots, often indicate the potential for such reconnection \citep{bakerTransientInverseFIPPlasma2019, bakerCanSubphotosphericMagnetic2020}.

Our estimation (see Section~\ref{sec_eventoveriew}) of the radial magnetic field, $\bm{B}_r$, being $\gtrsim$1300~G and $\gtrsim$900~G for flares FL1 and FL2, respectively, suggests that the equipartition layer resides either beneath the lower chromosphere or within the photosphere. This deduction is based on the fact that the equipartition layer is in the photosphere when $\bm{B}_r$ equals 300~G according to Laming's model.
The subchromospheric or subphotospheric reconnection generates upward sound waves, which become fast-mode MHD waves at the equipartition layer. The fast-mode waves undergo refraction or reflection in the chromosphere, turning to downward fast-mode waves.
These waves pose a downward ponderomotive force, pulling the low-FIP ions deeper and leading to the IFIP fractionation.
The strong magnetic field implies that the region situated in the upper chromosphere, where the FIP effect plasma usually resides, becomes thinner.
We can speculate that the duration of the IFIP effect is determined by the rate at which the FIP effect plasma is exhausted compared to the necessary fractionation rate to create FIP effect plasma.
Additionally, the decrease in elemental abundances during this period suggests reduced significance in the evaporation upflow speed and/or possibly a decline in the reflection of Alfvén waves. The latter is implied because the fractionation rate could be decreased by the reduced reflection of Alfvén waves.
In comparison, flare FL2 (Appendix Figure~\ref{fig_C1663AIA}), with a magnetic field of $\gtrsim$900~G, exhibits the case of plasma confined in closed magnetic loops, as opposed to flare FL1 (exhibiting the appearance of half-loops, see \citealp{guidoniTEMPERATUREELECTRONDENSITY2015}; see Section~\ref{sec_eventoveriew} and Appendix Figure~\ref{fig_M1365AIA}).
Note that the magnetic field for flare FL1 cannot be extrapolated because surface magnetic field measurements near the solar limb are not reliable.
We attribute the absence of the IFIP effect during the decay phase in this typical flare scenario to the continuous plasma replacement within the loop (i.e., reduced plasma replacement in FL1, as mentioned in Section~\ref{sec_decay} Decay phase) and to the wave reflection back and forth with the closed field configuration that increases the fractionation rate producing FIP effect plasma.
To fully understand why the IFIP effect is absent, further exploration of large eruptions, such as X-class flares with stronger magnetic fields and different morphologies, is required.

It is crucial to reiterate that this effect is particularly linked to the presence of highly complex magnetic structures during flares.
This makes it plausible that the potential subchromospheric or subphotospheric reconnection may principally take place (together with the other processes) during or even before flare eruption, and therefore persist for most of the flare lifetime.
This conjecture finds support from the lower FIP biases observed before the peak of flare FL1 (Figure~\ref{fig_M1365_FIP}), as compared to the averaged FIP biases over all flares reported by \citet{dennisSOLARFLAREELEMENT2015} (1.66 for Fe, 3.89 for Ca, 1.23 for S, 1.64 for Si, and 2.48 for Ar).
It is likely that the extent of this effect depends on the strength of the magnetic field.

Considering the origin of the IFIP effect plasma, it is natural to speculate that the IFIP effect for Fe could also be present but have a slower response due to its stronger inertia (see Section~\ref{sec_decay} Decay phase).
The transfer of Fe from the photosphere to the corona might require more time and could be measurable by CE-2/SXM if extended observations become feasible.
Additionally, it is reasonable to expect that the elemental abundances to recover to their pre-flare coronal levels. This recovery is similar to what is observed in most typical flare cases once the above-mentioned processes return to normal.

\section{Discussion and Conclusions}
\label{sec_summary}

\begin{figure}[ht!]%
	\centering
	\includegraphics[width=0.68\textwidth]{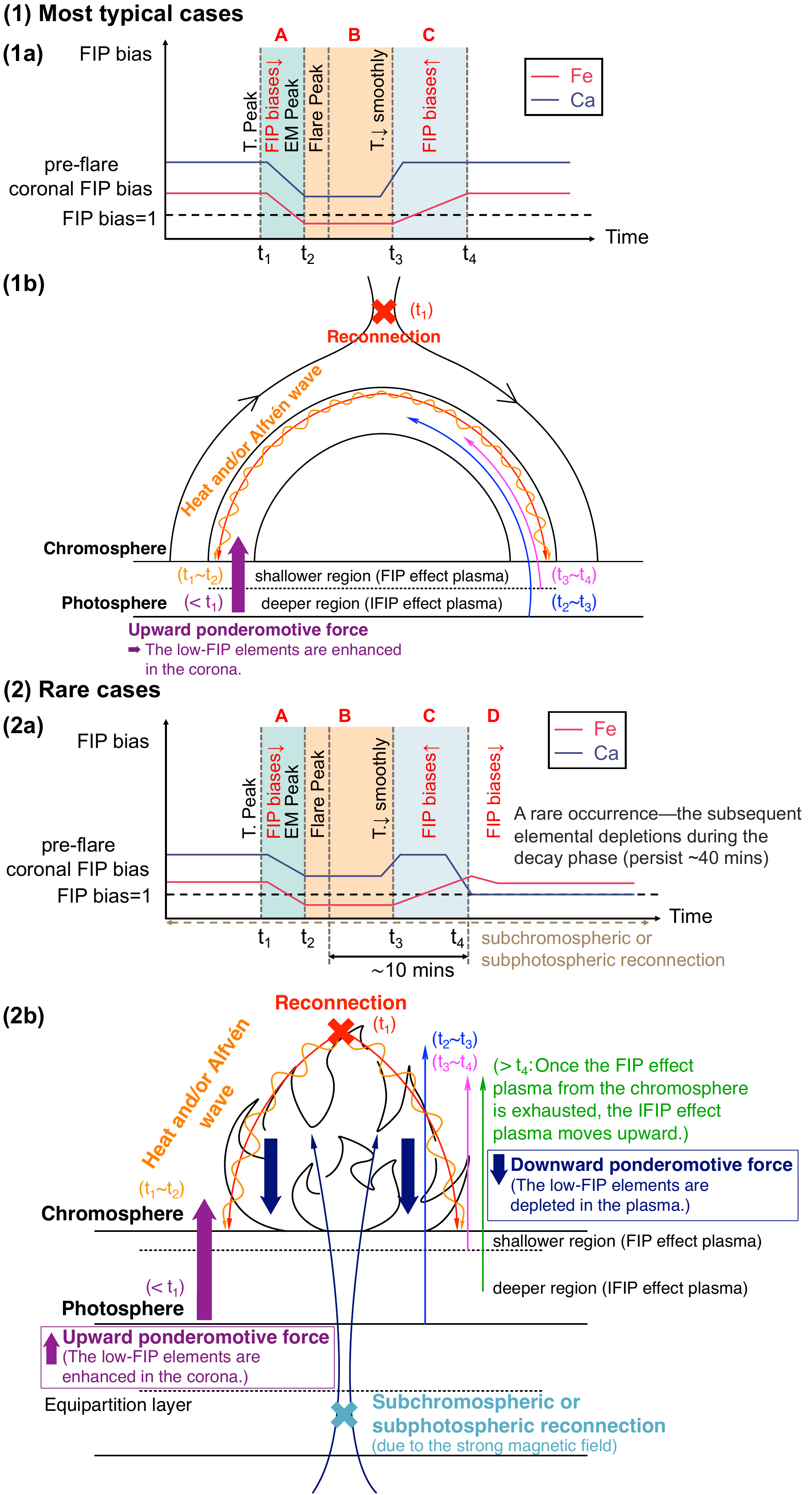}
	\caption{Schematic illustrations depict the evolution of elemental abundances during two flare scenarios: (1) most typical cases and (2) rare cases.
	Panels (1a) and (2a) depict the characterizations across phases: impulsive phase (A: $\sim{t_1}$--$t_2$; Section~\ref{sec_initial}), peak phase (B: $t_2$--$t_3$; Section~\ref{sec_peak}), and decay phase (C: $t_3$--$t_4$ and D: $>t_4$; Sections~\ref{sec_decay} and \ref{sec_subdecay}).
	Note that the flare peak represents the flux of a flare reaching its peak in soft X-ray range.
	Panels (1b) and (2b) visually present the physical processes occurring during flares.
	We remind readers to refer to Table~\ref{table_interpretation} for a concise summary of our physical interpretation (Section~\ref{sec_summary}), in conjunction with the information presented in the figure.
	The physical processes and mass transfers may occur at both footpoints, but for visual clarity, it is illustrated for one side.
	}
	\label{fig_interpretation}
\end{figure}

With the imaging and spectroscopic data analysis of the entire process in two solar flares, we propose a comprehensive interpretation of the abundance evolution expanding beyond the scope of previous studies that primarily focused on specific phases of the flare.
Our interpretation is summarized in the schematics presented in Figure~\ref{fig_interpretation}, shedding light on the temporal evolution of elemental abundances (Fe, Ca, S, Si, and Ar), temperature, and emission measure during solar flares in the active regions, ARs 11149 and 11158, with the observations of the Solar X-ray Monitor on-board the Chang'E-2 lunar orbiter.
In Figure~\ref{fig_interpretation}, panels (1a) and (2a) offer an overview of the characterization across different phases, illustrating the trend in the evolution of the FIP biases of the low-FIP elements (Fe and Ca), while panels (1b) and (2b) visually depict the physical processes occurring during flares.
For clarity, readers are reminded to refer to Table~\ref{table_interpretation}, which provides a summary of the depicted physical processes in Figure~\ref{fig_interpretation}.
Our observations during the peak phase revealed evident abundance depletion of the low-FIP elements (Ca, Fe, and Si), with Fe and Si exhibiting the IFIP effect by decreasing to their photospheric abundances.
Importantly, our results mark the discovery of the IFIP effect for Fe, and reaffirm its existence for Si.
During the decay phases, we observed a rapid recovery and return of elemental abundances to their pre-flare coronal levels.
However, an intriguing feature was revealed: heavier elements, particularly Fe, exhibited a slower recovery compared to all other lighter elements.
This discovery marks the first instance of such distinctive behavior, leading us to propose that the inertia effect plays a pivotal role in the evaporation process during flares.
Moreover, a rare phenomenon was disclosed during the decay phase of a large flare (FL1) showing second depletion of elemental abundances after the rapid return, suggesting its source from the lower chromosphere or even photosphere.
Our proposed explanation follows the CSHKP model to elucidate the formation of both the FIP and IFIP effects in the evolutionary dynamics of flares, with the ion inertia, in addition to the ponderomotive force, playing an important role.

\begin{table}
\begin{threeparttable}
\footnotesize
\caption{A concise summary of our interpretation, as visually illustrated in Figure~\ref{fig_interpretation}. The denotations are the same as those used in Figure~\ref{fig_interpretation}.}
\label{table_interpretation}
\newcolumntype{A}[1]{>{\raggedright\arraybackslash}p{#1}}
\newcolumntype{B}[1]{>{\raggedleft\arraybackslash}p{#1}}
\newcolumntype{C}[1]{>{\centering\arraybackslash}p{#1}}
\begin{tabularx}{\textwidth}{C{7em}A{4.5em}C{5em}A{40em}}
\toprule
Period & Time & FIP Biases & Physical Interpretation (in brief)\\
\midrule
Impulsive phase & A: $\sim$t$_1$--t$_2$ & $\downarrow$ & Rapid evaporation speed prevents FIP fractionation.\\
Peak phase & B: t$_2$--t$_3$ & --- & Evaporation continues to dominate over FIP fractionation. \newline Evaporation leads to the upward movement of plasma depleted of low-FIP elements (IFIP effect plasma) from the lower chromosphere.\\
Decay phase & C: t$_3$--t$_4$ & $\uparrow$ & Temperature decreases smoothly as the upflow speed decreases, with the dominant process gradually altering; in addition, the upward ponderomotive force generated by coronal Alfvén waves could contribute to the observed increase in the FIP biases. \newline The IFIP effect plasma is replaced with fractionated FIP effect plasma, with varying extents observed in (1) most typical cases and (2) rare cases. \\
\emph{ditto} \newline (rare cases) & D: $>$t$_4$ & $\downarrow$ and --- & Evaporation upflow speed becomes less significant and/or the reflection of Alfvén waves is nearly turned off. \newline Once the FIP effect plasma from the chromosphere is exhausted, the IFIP effect plasma moves upward.\\
\bottomrule
\end{tabularx}
\begin{tablenotes}
\item[] {\bf Note (for rare flare cases).} Subchromospheric or subphotospheric reconnection may occur during or even before flare eruption, persisting for most of the flare lifetime, together with the other processes.
\item[] {\bf Note.} The third column represents the trend in the evolution of the FIP biases of the \emph{low-FIP elements} (Ca, Fe, and Si), with $\uparrow$ denoting an increasing trend, $\downarrow$ representing a decreasing trend, and --- indicating no trend or a similar constant trend.
\end{tablenotes}
\end{threeparttable}
\end{table}

It is noted that the abundance distribution plot in \citet{dennisSOLARFLAREELEMENT2015} mentioned earlier (see Section~\ref{sec_intro}) implies that coronal abundances are thought to originate from the FIP fractionation processes that generate the abundances observed during flares. However, it is crucial to note that this argument is grounded in conventional understandings of the origin of coronal plasmas and theoretical modeling of the FIP effect, as described by \citet{lamingUnifiedPictureFirst2004}.
The origin of coronal plasmas may involve several distinct physical mechanisms.
As a flare evolves, so does the FIP bias of elemental abundances.
Therefore, averaging the FIP biases over a large number of flares but within small time window, as carried out in the statistical analysis of 526 large flares by \citet{dennisSOLARFLAREELEMENT2015}, provides an average value that involves the various mechanisms at play from flare to flare.
Specifically, the continually changing FIP bias of elemental abundances in a flare likely reflects a combination of the FIP effect plasma and a certain extent of IFIP effect plasma.
Therefore, we argue instead that these averaged abundances are unlikely to give an accurate estimation of the true coronal abundances, nor do they fully account for the FIP fractionation occurring in flares for elements with FIPs of $\lesssim$7~eV. This is because the underlying mechanism of flare plasma formation is not solely due to FIP fractionation.

Besides, \citet{toEvolutionPlasmaComposition2021} reported the inconsistent behavior of the \ion{Ca}{14}/\ion{Ar}{14} intensity ratio within the X-shaped structure of a very small flare (AR 11967) that was barely identified in the GOES X-ray light curve.
Based on the observations from Hinode/EIS, this behavior indicates a significant difference in the Ca/Ar ratio values of $\sim$2 and 3.77 (with a typical value of $\sim$2) between lower and higher temperatures. This difference suggests that the higher intensity ratio, representing a stronger FIP effect, is observed at higher temperature parts in smaller flares.
Therefore, the differing flare structures may lead to contrasting behavior related to temperature, possibly indicating that various processes are at play for different flare structures, which requires further investigations.

\begin{acknowledgments}
This research was supported by the Science and Technology Development Fund (FDCT) of Macau (grant Nos. 0014/2022/A1, SKL-LPS(MUST)-2021-2023) and NSFC (12127901).
Scientific data of Chang'E-2 mission are provided by the China National Space Administration (CNSA). We are grateful for the support from the team members of the Ground Research and Application System (GRAS), who contributed to data receiving and preprocessing.
We thank the teams of SDO and GOES for providing the excellent data.
We thank the anonymous reviewer for providing us with the insightful and helpful suggestions and comments.
We acknowledge Brian Dennis, Richard Schwartz, and Kim Tolbert at NASA/Goddard Space Flight Center (NASA/GSFC) for the valuable discussions, making an initial DRM suitable for CE-2/SXM, and assistance with OSPEX, respectively.
\end{acknowledgments}

\software{OSPEX~\citep{tolbertOSPEXObjectSpectral2020}, Astropy~\citep{theastropycollaborationAstropyProjectSustaining2022}, Sunpy~\citep{thesunpycommunitySunPyProjectInteroperable2023}, CHIANTI~\citep{dereCHIANTIAtomicDatabase1997, dereCHIANTIAtomicDatabase2019}}

\bibliographystyle{aasjournal}

\begin{thebibliography}{}
\expandafter\ifx\csname natexlab\endcsname\relax\def\natexlab#1{#1}\fi
\providecommand{\url}[1]{\href{#1}{#1}}
\providecommand{\dodoi}[1]{doi:~\href{http://doi.org/#1}{\nolinkurl{#1}}}
\providecommand{\doeprint}[1]{\href{http://ascl.net/#1}{\nolinkurl{http://ascl.net/#1}}}
\providecommand{\doarXiv}[1]{\href{https://arxiv.org/abs/#1}{\nolinkurl{https://arxiv.org/abs/#1}}}

\bibitem[{Asplund {et~al.}(2009)Asplund, Grevesse, Sauval, \&
  Scott}]{asplundChemicalCompositionSun2009}
Asplund, M., Grevesse, N., Sauval, A.~J., \& Scott, P. 2009, Annual Review of
  Astronomy and Astrophysics, 47, 481,
  \dodoi{10.1146/annurev.astro.46.060407.145222}

\bibitem[{Aulanier {et~al.}(2013)Aulanier, D{\'e}moulin, Schrijver, Janvier,
  Pariat, \& Schmieder}]{aulanierStandardFlareModel2013}
Aulanier, G., D{\'e}moulin, P., Schrijver, C.~J., {et~al.} 2013, Astronomy \&
  Astrophysics, 549, A66, \dodoi{10.1051/0004-6361/201220406}

\bibitem[{Baker {et~al.}(2013)Baker, Brooks, D{\'e}moulin, {van
  Driel-Gesztelyi}, Green, Steed, \&
  Carlyle}]{bakerPlasmaCompositionSigmoidal2013}
Baker, D., Brooks, D.~H., D{\'e}moulin, P., {et~al.} 2013, The Astrophysical
  Journal, 778, 69, \dodoi{10.1088/0004-637X/778/1/69}

\bibitem[{Baker {et~al.}(2019)Baker, van {Driel-Gesztelyi}, Brooks, Valori,
  James, Laming, Long, D{\'e}moulin, Green, Matthews, Ol{\'a}h, \& K{\H
  o}v{\'a}ri}]{bakerTransientInverseFIPPlasma2019}
Baker, D., van {Driel-Gesztelyi}, L., Brooks, D.~H., {et~al.} 2019, The
  Astrophysical Journal, 875, 35, \dodoi{10.3847/1538-4357/ab07c1}

\bibitem[{Baker {et~al.}(2020)Baker, {van Driel-Gesztelyi}, Brooks,
  D{\'e}moulin, Valori, Long, Laming, To, \&
  James}]{bakerCanSubphotosphericMagnetic2020}
Baker, D., {van Driel-Gesztelyi}, L., Brooks, D.~H., {et~al.} 2020, The
  Astrophysical Journal, 894, 35, \dodoi{10.3847/1538-4357/ab7dcb}

\bibitem[{Ban {et~al.}(2014)Ban, Zheng, Zhu, Zhang, Xu, \&
  Zou}]{banResearchInversionElemental2014a}
Ban, C., Zheng, Y., Zhu, Y., {et~al.} 2014, Chinese Journal of Geochemistry,
  33, 289, \dodoi{10.1007/s11631-014-0690-2}

\bibitem[{Benz(2008)}]{benzFlareObservations2008}
Benz, A.~O. 2008, Living Reviews in Solar Physics, 5, 1,
  \dodoi{10.12942/lrsp-2008-1}

\bibitem[{Bobra {et~al.}(2014)Bobra, Sun, Hoeksema, Turmon, Liu, Hayashi,
  Barnes, \& Leka}]{bobraHelioseismicMagneticImager2014}
Bobra, M.~G., Sun, X., Hoeksema, J.~T., {et~al.} 2014, Solar Physics, 289,
  3549, \dodoi{10.1007/s11207-014-0529-3}

\bibitem[{Botha {et~al.}(2011)Botha, Arber, \&
  Hood}]{bothaThermalConductionEffects2011}
Botha, G. J.~J., Arber, T.~D., \& Hood, A.~W. 2011, Astronomy \& Astrophysics,
  525, A96, \dodoi{10.1051/0004-6361/201015534}

\bibitem[{Brooks(2018)}]{brooksDiagnosticCoronalElemental2018}
Brooks, D.~H. 2018, The Astrophysical Journal, 863, 140,
  \dodoi{10.3847/1538-4357/aad415}

\bibitem[{Cargill(2014)}]{cargillActiveRegionEmission2014}
Cargill, P.~J. 2014, The Astrophysical Journal, 784, 49,
  \dodoi{10.1088/0004-637X/784/1/49}

\bibitem[{Cargill \& Klimchuk(2004)}]{cargillNanoflareHeatingCorona2004}
Cargill, P.~J., \& Klimchuk, J.~A. 2004, The Astrophysical Journal, 605, 911,
  \dodoi{10.1086/382526}

\bibitem[{Carmichael(1964)}]{carmichaelProcessFlares1964}
Carmichael, H. 1964, NASA Special Publication, 50, 451

\bibitem[{Caspi {et~al.}(2015)Caspi, Woods, \&
  Warren}]{caspiNEWOBSERVATIONSSOLAR2015}
Caspi, A., Woods, T.~N., \& Warren, H.~P. 2015, The Astrophysical Journal
  Letters, 802, L2, \dodoi{10.1088/2041-8205/802/1/L2}

\bibitem[{Chen(2011)}]{chenCoronalMassEjections2011}
Chen, P.~F. 2011, Living Reviews in Solar Physics, 8, 1,
  \dodoi{10.12942/lrsp-2011-1}

\bibitem[{Culhane {et~al.}(2007)Culhane, Harra, James, {Al-Janabi}, Bradley,
  Chaudry, Rees, Tandy, Thomas, Whillock, Winter, Doschek, Korendyke, Brown,
  Myers, Mariska, Seely, Lang, Kent, Shaughnessy, Young, Simnett, Castelli,
  Mahmoud, {Mapson-Menard}, Probyn, Thomas, Davila, Dere, Windt, Shea, Hagood,
  Moye, Hara, Watanabe, Matsuzaki, Kosugi, Hansteen, \&
  Wikstol}]{culhaneEUVImagingSpectrometer2007}
Culhane, J.~L., Harra, L.~K., James, A.~M., {et~al.} 2007, Solar Physics, 243,
  19, \dodoi{10.1007/s01007-007-0293-1}

\bibitem[{Dennis {et~al.}(2015)Dennis, Phillips, Schwartz, Tolbert, Starr, \&
  Nittler}]{dennisSOLARFLAREELEMENT2015}
Dennis, B.~R., Phillips, K. J.~H., Schwartz, R.~A., {et~al.} 2015, The
  Astrophysical Journal, 803, 67, \dodoi{10.1088/0004-637X/803/2/67}

\bibitem[{Dere {et~al.}(2019)Dere, Del~Zanna, Young, Landi, \&
  Sutherland}]{dereCHIANTIAtomicDatabase2019}
Dere, K.~P., Del~Zanna, G., Young, P.~R., Landi, E., \& Sutherland, R.~S. 2019,
  The Astrophysical Journal Supplement Series, 241, 22,
  \dodoi{10.3847/1538-4365/ab05cf}

\bibitem[{Dere {et~al.}(1997)Dere, Landi, Mason, Monsignori~Fossi, \&
  Young}]{dereCHIANTIAtomicDatabase1997}
Dere, K.~P., Landi, E., Mason, H.~E., Monsignori~Fossi, B.~C., \& Young, P.~R.
  1997, Astronomy and Astrophysics Supplement Series, 125, 149,
  \dodoi{10.1051/aas:1997368}

\bibitem[{Dong {et~al.}(2019)Dong, Zhang, Li, Tang, Xu, \&
  Zhang}]{dongCalibrationsChangE22019}
Dong, W.-D., Zhang, X., Li, Y., {et~al.} 2019, Solar Physics, 294, 120,
  \dodoi{10.1007/s11207-019-1508-5}

\bibitem[{Doschek \& Warren(2016)}]{doschekMYSTERIOUSCASESOLAR2016}
Doschek, G.~A., \& Warren, H.~P. 2016, The Astrophysical Journal, 825, 36,
  \dodoi{10.3847/0004-637X/825/1/36}

\bibitem[{Doschek \& Warren(2017)}]{doschekSunspotsStarspotsElemental2017}
---. 2017, The Astrophysical Journal, 844, 52, \dodoi{10.3847/1538-4357/aa7bea}

\bibitem[{Doschek {et~al.}(2015)Doschek, Warren, \&
  Feldman}]{doschekAnomalousRelativeAr2015}
Doschek, G.~A., Warren, H.~P., \& Feldman, U. 2015, The Astrophysical Journal,
  808, L7, \dodoi{10.1088/2041-8205/808/1/L7}

\bibitem[{Doschek {et~al.}(2018)Doschek, Warren, Harra, Culhane, Watanabe, \&
  Hara}]{doschekPhotosphericCoronalAbundances2018}
Doschek, G.~A., Warren, H.~P., Harra, L.~K., {et~al.} 2018, The Astrophysical
  Journal, 853, 178, \dodoi{10.3847/1538-4357/aaa4f5}

\bibitem[{Feldman(1992)}]{feldmanElementalAbundancesUpper1992}
Feldman, U. 1992, Physica Scripta, 46, 202, \dodoi{10.1088/0031-8949/46/3/002}

\bibitem[{Feldman \& Laming(2000)}]{feldmanElementAbundancesUpper2000}
Feldman, U., \& Laming, J.~M. 2000, Physica Scripta, 61, 222,
  \dodoi{10.1238/Physica.Regular.061a00222}

\bibitem[{Fisher {et~al.}(1985)Fisher, Canfield, \&
  McClymont}]{fisherFlareLoopRadiative1985}
Fisher, G.~H., Canfield, R.~C., \& McClymont, A.~N. 1985, The Astrophysical
  Journal, 289, 425, \dodoi{10.1086/162902}

\bibitem[{Fletcher \& Hudson(2008)}]{fletcherImpulsivePhaseFlare2008}
Fletcher, L., \& Hudson, H.~S. 2008, The Astrophysical Journal, 675, 1645,
  \dodoi{10.1086/527044}

\bibitem[{Fludra \& Schmelz(1999)}]{fludraAbsoluteCoronalAbundances1999}
Fludra, A., \& Schmelz, J.~T. 1999, Astronomy and Astrophysics, 348, 286

\bibitem[{Forbes \& Acton(1996)}]{forbesReconnectionFieldLine1996}
Forbes, T.~G., \& Acton, L.~W. 1996, The Astrophysical Journal, 459, 330,
  \dodoi{10.1086/176896}

\bibitem[{Forbes {et~al.}(1989)Forbes, Malherbe, \&
  Priest}]{forbesFormationFlareLoops1989}
Forbes, T.~G., Malherbe, J.~M., \& Priest, E.~R. 1989, Solar Physics, 120, 285,
  \dodoi{10.1007/BF00159881}

\bibitem[{Freeland \& Handy(1998)}]{freelandDataAnalysisSolarSoft1998}
Freeland, S.~L., \& Handy, B.~N. 1998, Solar Physics, 182, 497,
  \dodoi{10.1023/A:1005038224881}

\bibitem[{Gold {et~al.}(2001)Gold, Solomon, McNutt, Santo, Abshire, Acu{\~n}a,
  Afzal, Anderson, Andrews, Bedini, Cain, Cheng, Evans, Feldman, Follas,
  Gloeckler, Goldsten, Hawkins~III, Izenberg, Jaskulek, Ketchum, Lankton, Lohr,
  Mauk, McClintock, Murchie, Schlemm~II, Smith, Starr, \&
  Zurbuchen}]{goldMESSENGERMissionMercury2001}
Gold, R.~E., Solomon, S.~C., McNutt, R.~L., {et~al.} 2001, Planetary and Space
  Science, 49, 1467, \dodoi{10.1016/S0032-0633(01)00086-1}

\bibitem[{Guidoni {et~al.}(2015)Guidoni, McKenzie, Longcope, Plowman, \&
  Yoshimura}]{guidoniTEMPERATUREELECTRONDENSITY2015}
Guidoni, S.~E., McKenzie, D.~E., Longcope, D.~W., Plowman, J.~E., \& Yoshimura,
  K. 2015, The Astrophysical Journal, 800, 54,
  \dodoi{10.1088/0004-637X/800/1/54}

\bibitem[{Hirayama(1974)}]{hirayamaTheoreticalModelFlares1974}
Hirayama, T. 1974, Solar Physics, 34, 323, \dodoi{10.1007/BF00153671}

\bibitem[{Katsuda {et~al.}(2020)Katsuda, Ohno, Mori, Beppu, Kanemaru, Tashiro,
  Terada, Sato, Morita, Sagara, Ogawa, Takahashi, Murakami, Nobukawa, Tsunemi,
  Hayashida, Matsumoto, Noda, Nakajima, Ezoe, Tsuboi, Maeda, Yokoyama, \&
  Narukage}]{katsudaInverseFirstIonization2020}
Katsuda, S., Ohno, M., Mori, K., {et~al.} 2020, The Astrophysical Journal, 891,
  126, \dodoi{10.3847/1538-4357/ab7207}

\bibitem[{Klimchuk(2006)}]{klimchukSolvingCoronalHeating2006}
Klimchuk, J.~A. 2006, Solar Physics, 234, 41, \dodoi{10.1007/s11207-006-0055-z}

\bibitem[{Klimchuk(2017)}]{klimchukNanoflareHeatingObservations2017}
---. 2017, Nanoflare {{Heating}}: {{Observations}} and {{Theory}},
  \dodoi{10.48550/arXiv.1709.07320}

\bibitem[{Kopp \& Pneuman(1976)}]{koppMagneticReconnectionCorona1976}
Kopp, R.~A., \& Pneuman, G.~W. 1976, Solar Physics, 50, 85,
  \dodoi{10.1007/BF00206193}

\bibitem[{Kosugi {et~al.}(2007)Kosugi, Matsuzaki, Sakao, Shimizu, Sone,
  Tachikawa, Hashimoto, Minesugi, Ohnishi, Yamada, Tsuneta, Hara, Ichimoto,
  Suematsu, Shimojo, Watanabe, Shimada, Davis, Hill, Owens, Title, Culhane,
  Harra, Doschek, \& Golub}]{kosugiHinodeSolarBMission2007}
Kosugi, T., Matsuzaki, K., Sakao, T., {et~al.} 2007, Solar Physics, 243, 3,
  \dodoi{10.1007/s11207-007-9014-6}

\bibitem[{Koyama {et~al.}(2007)Koyama, Tsunemi, Dotani, Bautz, Hayashida,
  Tsuru, Matsumoto, Ogawara, Ricker, Doty, Kissel, Foster, Nakajima, Yamaguchi,
  Mori, Sakano, Hamaguchi, Nishiuchi, Miyata, Torii, Namiki, Katsuda, Matsuura,
  Miyauchi, Anabuki, Tawa, Ozaki, Murakami, Maeda, Ichikawa, Prigozhin,
  Boughan, Lamarr, Miller, Burke, Gregory, Pillsbury, Bamba, Hiraga, Senda,
  Katayama, Kitamoto, Tsujimoto, Kohmura, Tsuboi, \&
  Awaki}]{koyamaXRayImagingSpectrometer2007}
Koyama, K., Tsunemi, H., Dotani, T., {et~al.} 2007, Publications of the
  Astronomical Society of Japan, 59, 23, \dodoi{10.1093/pasj/59.sp1.S23}

\bibitem[{Laming(2004)}]{lamingUnifiedPictureFirst2004}
Laming, J.~M. 2004, The Astrophysical Journal, 614, 1063,
  \dodoi{10.1086/423780}

\bibitem[{Laming(2009)}]{lamingNONWKBMODELSFIRST2009}
---. 2009, The Astrophysical Journal, 695, 954,
  \dodoi{10.1088/0004-637X/695/2/954}

\bibitem[{Laming(2012)}]{lamingNONWKBMODELSFIRST2012}
---. 2012, The Astrophysical Journal, 744, 115,
  \dodoi{10.1088/0004-637X/744/2/115}

\bibitem[{Laming(2015)}]{lamingFIPInverseFIP2015}
---. 2015, Living Reviews in Solar Physics, 12, 2, \dodoi{10.1007/lrsp-2015-2}

\bibitem[{Laming(2017)}]{lamingFirstIonizationPotential2017}
---. 2017, The Astrophysical Journal, 844, 153,
  \dodoi{10.3847/1538-4357/aa7cf1}

\bibitem[{Laming(2021)}]{lamingFIPInverseFIPEffects2021}
---. 2021, The Astrophysical Journal, 909, 17, \dodoi{10.3847/1538-4357/abd9c3}

\bibitem[{Laming \& Hwang(2009)}]{lamingTHERMALCONDUCTIVITYELEMENT2009}
Laming, J.~M., \& Hwang, U. 2009, The Astrophysical Journal, 707, L60,
  \dodoi{10.1088/0004-637X/707/1/L60}

\bibitem[{Laming {et~al.}(2019)Laming, Vourlidas, Korendyke, Chua, Cranmer, Ko,
  Kuroda, Provornikova, Raymond, Raouafi, Strachan, {Tun-Beltran}, Weberg, \&
  Wood}]{lamingElementAbundancesNew2019}
Laming, J.~M., Vourlidas, A., Korendyke, C., {et~al.} 2019, The Astrophysical
  Journal, 879, 124, \dodoi{10.3847/1538-4357/ab23f1}

\bibitem[{Lang(2006)}]{langSunEarthSky2006}
Lang, K.~R., ed. 2006, Sun, {{Earth}} and {{Sky}}, 2nd edn. (New York, NY:
  Springer), \dodoi{10.1007/978-0-387-33365-6}

\bibitem[{Lemen {et~al.}(2012)Lemen, Title, Akin, Boerner, Chou, Drake, Duncan,
  Edwards, Friedlaender, Heyman, Hurlburt, Katz, Kushner, Levay, Lindgren,
  Mathur, McFeaters, Mitchell, Rehse, Schrijver, Springer, Stern, Tarbell,
  Wuelser, Wolfson, Yanari, Bookbinder, Cheimets, Caldwell, Deluca, Gates,
  Golub, Park, Podgorski, Bush, Scherrer, Gummin, Smith, Auker, Jerram, Pool,
  Soufli, Windt, Beardsley, Clapp, Lang, \&
  Waltham}]{lemenAtmosphericImagingAssembly2012}
Lemen, J.~R., Title, A.~M., Akin, D.~J., {et~al.} 2012, Solar Physics, 275, 17,
  \dodoi{10.1007/s11207-011-9776-8}

\bibitem[{Lin \& Hudson(1976)}]{linNonthermalProcessesLarge1976}
Lin, R.~P., \& Hudson, H.~S. 1976, Solar Physics, 50, 153,
  \dodoi{10.1007/BF00206199}

\bibitem[{Mazzotta {et~al.}(1998)Mazzotta, Mazzitelli, Colafrancesco, \&
  Vittorio}]{mazzottaIonizationBalanceOptically1998}
Mazzotta, P., Mazzitelli, G., Colafrancesco, S., \& Vittorio, N. 1998,
  Astronomy and Astrophysics Supplement Series, 133, 403,
  \dodoi{10.1051/aas:1998330}

\bibitem[{Meyer(1985)}]{meyerSolarstellarOuterAtmospheres1985}
Meyer, J.~P. 1985, The Astrophysical Journal Supplement Series, 57, 173,
  \dodoi{10.1086/191001}

\bibitem[{Mitsuda {et~al.}(2007)Mitsuda, Bautz, Inoue, Kelley, Koyama, Kunieda,
  Makishima, Ogawara, Petre, Takahashi, Tsunemi, White, Anabuki, Angelini,
  Arnaud, Awaki, Bamba, Boyce, Brown, Chan, Cottam, Dotani, Doty, Ebisawa,
  Ezoe, Fabian, Figueroa, Fujimoto, Fukazawa, Furusho, Furuzawa, Gendreau,
  Griffiths, Haba, Hamaguchi, Harrus, Hasinger, Hatsukade, Hayashida, Henry,
  Hiraga, Holt, Hornschemeier, Hughes, Hwang, Ishida, Ishisaki, Isobe, Itoh,
  Iyomoto, Kahn, Kamae, Katagiri, Kataoka, Katayama, Kawai, Kilbourne,
  Kinugasa, Kissel, Kitamoto, Kohama, Kohmura, Kokubun, Kotani, Kotoku, Kubota,
  Madejski, Maeda, Makino, Markowitz, Matsumoto, Matsumoto, Matsuoka,
  Matsushita, McCammon, Mihara, Misaki, Miyata, Mizuno, Mori, Mori, Morii,
  Moseley, Mukai, Murakami, Murakami, Mushotzky, Nagase, Namiki, Negoro,
  Nakazawa, Nousek, Okajima, Ogasaka, Ohashi, Oshima, Ota, Ozaki, Ozawa,
  Parmar, Pence, Porter, Reeves, Ricker, Sakurai, Sanders, Senda, Serlemitsos,
  Shibata, Soong, Smith, Suzuki, Szymkowiak, Takahashi, Tamagawa, Tamura,
  Tamura, Tanaka, Tashiro, Tawara, Terada, Terashima, Tomida, Torii, Tsuboi,
  Tsujimoto, Tsuru, Turner, Ueda, Ueno, Ueno, Uno, Urata, Watanabe, Yamamoto,
  Yamaoka, Yamasaki, Yamashita, Yamauchi, Yamauchi, Yaqoob, Yonetoku, \&
  Yoshida}]{mitsudaXRayObservatorySuzaku2007}
Mitsuda, K., Bautz, M., Inoue, H., {et~al.} 2007, Publications of the
  Astronomical Society of Japan, 59, S1, \dodoi{10.1093/pasj/59.sp1.S1}

\bibitem[{Mondal {et~al.}(2021)Mondal, Sarkar, Vadawale, Mithun, Janardhan,
  Zanna, Mason, {Mitra-Kraev}, \&
  Narendranath}]{mondalEvolutionElementalAbundances2021}
Mondal, B., Sarkar, A., Vadawale, S.~V., {et~al.} 2021, The Astrophysical
  Journal, 920, 4, \dodoi{10.3847/1538-4357/ac14c1}

\bibitem[{Nama {et~al.}(2023)Nama, Mondal, Narendranath, \&
  Paul}]{namaCoronalElementalAbundances2023}
Nama, L., Mondal, B., Narendranath, S., \& Paul, K.~T. 2023, Solar Physics,
  298, 55, \dodoi{10.1007/s11207-023-02142-5}

\bibitem[{Narendranath {et~al.}(2014)Narendranath, Sreekumar, Alha,
  Sankarasubramanian, Huovelin, \&
  Athiray}]{narendranathElementalAbundancesSolar2014}
Narendranath, S., Sreekumar, P., Alha, L., {et~al.} 2014, Solar Physics, 289,
  1585, \dodoi{10.1007/s11207-013-0410-9}

\bibitem[{Narendranath {et~al.}(2020)Narendranath, Sreekumar, Pillai, Panini,
  Sankarasubramanian, \& Huovelin}]{narendranathCoronalElementalAbundance2020}
Narendranath, S., Sreekumar, P., Pillai, N.~S., {et~al.} 2020, Solar Physics,
  295, 175, \dodoi{10.1007/s11207-020-01738-5}

\bibitem[{Neupert(1968)}]{neupertComparisonSolarXRay1968}
Neupert, W.~M. 1968, The Astrophysical Journal, 153, L59,
  \dodoi{10.1086/180220}

\bibitem[{Ning \& Cao(2010)}]{ningINVESTIGATIONCHROMOSPHERICEVAPORATION2010}
Ning, Z., \& Cao, W. 2010, The Astrophysical Journal, 717, 1232,
  \dodoi{10.1088/0004-637X/717/2/1232}

\bibitem[{Parker(1972)}]{parkerTopologicalDissipationSmallScale1972}
Parker, E.~N. 1972, The Astrophysical Journal, 174, 499, \dodoi{10.1086/151512}

\bibitem[{Peng {et~al.}(2009)Peng, Wang, Zhang, Cui, Cao, Zhang, Liang, Wang,
  Gao, Yang, \& Wu}]{pengProspectiveResultsCHANG2009}
Peng, W.-X., Wang, H.-Y., Zhang, C.-M., {et~al.} 2009, Chinese Physics C, 33,
  819, \dodoi{10.1088/1674-1137/33/10/001}

\bibitem[{Pesnell {et~al.}(2012)Pesnell, Thompson, \&
  Chamberlin}]{pesnellSolarDynamicsObservatory2012}
Pesnell, W.~D., Thompson, B.~J., \& Chamberlin, P.~C. 2012, Solar Physics, 275,
  3, \dodoi{10.1007/s11207-011-9841-3}

\bibitem[{Phillips \& Dennis(2012)}]{phillipsSOLARFLAREIRON2012}
Phillips, K. J.~H., \& Dennis, B.~R. 2012, The Astrophysical Journal, 748, 52,
  \dodoi{10.1088/0004-637X/748/1/52}

\bibitem[{Priest \& Forbes(2002)}]{priestMagneticNatureSolar2002}
Priest, E., \& Forbes, T. 2002, The Astronomy and Astrophysics Review, 10, 313,
  \dodoi{10.1007/s001590100013}

\bibitem[{Reale {et~al.}(2016)Reale, Orlando, Guarrasi, Mignone, Peres, Hood,
  \& Priest}]{reale3DMHDMODELING2016}
Reale, F., Orlando, S., Guarrasi, M., {et~al.} 2016, The Astrophysical Journal,
  830, 21, \dodoi{10.3847/0004-637X/830/1/21}

\bibitem[{Saba(1995)}]{sabaSpectroscopicMeasurementsElement1995}
Saba, J. L.~R. 1995, Advances in Space Research, 15, 13

\bibitem[{Santo {et~al.}(2001)Santo, Gold, McNutt, Solomon, Ercol, Farquhar,
  Hartka, Jenkins, McAdams, Mosher, Persons, Artis, Bokulic, Conde, Dakermanji,
  Goss, Haley, Heeres, Maurer, Moore, Rodberg, Stern, Wiley, Williams, Yen, \&
  Peterson}]{santoMESSENGERMissionMercury2001}
Santo, A.~G., Gold, R.~E., McNutt, R.~L., {et~al.} 2001, Planetary and Space
  Science, 49, 1481, \dodoi{10.1016/S0032-0633(01)00087-3}

\bibitem[{Scherrer {et~al.}(2012)Scherrer, Schou, Bush, Kosovichev, Bogart,
  Hoeksema, Liu, Duvall, Zhao, Title, Schrijver, Tarbell, \&
  Tomczyk}]{scherrerHelioseismicMagneticImager2012}
Scherrer, P.~H., Schou, J., Bush, R.~I., {et~al.} 2012, Solar Physics, 275,
  207, \dodoi{10.1007/s11207-011-9834-2}

\bibitem[{Schlemm {et~al.}(2007)Schlemm, Starr, Ho, Bechtold, Hamilton, Boldt,
  Boynton, Bradley, Fraeman, Gold, Goldsten, Hayes, Jaskulek, Rossano, Rumpf,
  Schaefer, Strohbehn, Shelton, Thompson, Trombka, \&
  Williams}]{schlemmXRaySpectrometerMESSENGER2007}
Schlemm, C.~E., Starr, R.~D., Ho, G.~C., {et~al.} 2007, Space Science Reviews,
  131, 393, \dodoi{10.1007/s11214-007-9248-5}

\bibitem[{Schmelz {et~al.}(2012)Schmelz, Reames, {von Steiger}, \&
  Basu}]{schmelzCOMPOSITIONSOLARCORONA2012}
Schmelz, J.~T., Reames, D.~V., {von Steiger}, R., \& Basu, S. 2012, The
  Astrophysical Journal, 755, 33, \dodoi{10.1088/0004-637X/755/1/33}

\bibitem[{Schou {et~al.}(2012)Schou, Scherrer, Bush, Wachter, Couvidat,
  {Rabello-Soares}, Bogart, Hoeksema, Liu, Duvall, Akin, Allard, Miles,
  Rairden, Shine, Tarbell, Title, Wolfson, Elmore, Norton, \&
  Tomczyk}]{schouDesignGroundCalibration2012}
Schou, J., Scherrer, P.~H., Bush, R.~I., {et~al.} 2012, Solar Physics, 275,
  229, \dodoi{10.1007/s11207-011-9842-2}

\bibitem[{Shibata \& Magara(2011)}]{shibataSolarFlaresMagnetohydrodynamic2011}
Shibata, K., \& Magara, T. 2011, Living Reviews in Solar Physics, 8,
  \dodoi{10.12942/lrsp-2011-6}

\bibitem[{Solomon {et~al.}(2001)Solomon, McNutt, Gold, Acu{\~n}a, Baker,
  Boynton, Chapman, Cheng, Gloeckler, Head~III, Krimigis, McClintock, Murchie,
  Peale, Phillips, Robinson, Slavin, Smith, Strom, Trombka, \&
  Zuber}]{solomonMESSENGERMissionMercury2001}
Solomon, S.~C., McNutt, R.~L., Gold, R.~E., {et~al.} 2001, Planetary and Space
  Science, 49, 1445, \dodoi{10.1016/S0032-0633(01)00085-X}

\bibitem[{Sturrock(1966)}]{sturrockModelHighEnergyPhase1966}
Sturrock, P.~A. 1966, Nature, 211, 695, \dodoi{10.1038/211695a0}

\bibitem[{Sylwester {et~al.}(2014)Sylwester, Sylwester, Phillips, K{\k e}pa, \&
  Mrozek}]{sylwesterSolarFlareComposition2014}
Sylwester, B., Sylwester, J., Phillips, K. J.~H., K{\k e}pa, A., \& Mrozek, T.
  2014, The Astrophysical Journal, 787, 122,
  \dodoi{10.1088/0004-637X/787/2/122}

\bibitem[{Tam(2014)}]{tamMHDSimulationsCoronal2014}
Tam, K.~V. 2014, Thesis, University of St Andrews

\bibitem[{{The Astropy Collaboration} {et~al.}(2022){The Astropy
  Collaboration}, {Price-Whelan}, Lim, Earl, Starkman, Bradley, Shupe, Patil,
  Corrales, Brasseur, N{\"o}the, Donath, Tollerud, Morris, Ginsburg, Vaher,
  Weaver, Tocknell, Jamieson, van Kerkwijk, Robitaille, Merry, Bachetti,
  G{\"u}nther, Authors, Aldcroft, {Alvarado-Montes}, Archibald, B{\'o}di,
  Bapat, Barentsen, Baz{\'a}n, Biswas, Boquien, Burke, Cara, Cara, Conroy,
  Conseil, Craig, Cross, Cruz, D'Eugenio, Dencheva, Devillepoix, Dietrich,
  Eigenbrot, Erben, Ferreira, {Foreman-Mackey}, Fox, Freij, Garg, Geda,
  Glattly, Gondhalekar, Gordon, Grant, Greenfield, Groener, Guest, Gurovich,
  Handberg, Hart, {Hatfield-Dodds}, Homeier, Hosseinzadeh, Jenness, Jones,
  Joseph, Kalmbach, Karamehmetoglu, Ka{\l}uszy{\'n}ski, Kelley, Kern,
  Kerzendorf, Koch, Kulumani, Lee, Ly, Ma, MacBride, Maljaars, Muna, Murphy,
  Norman, O'Steen, Oman, Pacifici, Pascual, {Pascual-Granado}, Patil, Perren,
  Pickering, Rastogi, Roulston, Ryan, Rykoff, Sabater, Sakurikar, Salgado,
  Sanghi, Saunders, Savchenko, Schwardt, {Seifert-Eckert}, Shih, Jain, Shukla,
  Sick, Simpson, Singanamalla, Singer, Singhal, Sinha, Sip{\H o}cz, Spitler,
  Stansby, Streicher, {\v S}umak, Swinbank, Taranu, Tewary, Tremblay,
  de~{Val-Borro}, Kooten, Vasovi{\'c}, Verma, Cardoso, Williams, Wilson,
  Winkel, {Wood-Vasey}, Xue, Yoachim, Zhang, Zonca, \&
  Contributors}]{theastropycollaborationAstropyProjectSustaining2022}
{The Astropy Collaboration}, {Price-Whelan}, A.~M., Lim, P.~L., {et~al.} 2022,
  The Astrophysical Journal, 935, 167, \dodoi{10.3847/1538-4357/ac7c74}

\bibitem[{{The SunPy Community} {et~al.}(2023){The SunPy Community}, Barnes,
  Christe, Freij, Hayes, Stansby, Ireland, Mumford, Ryan, \&
  Shih}]{thesunpycommunitySunPyProjectInteroperable2023}
{The SunPy Community}, Barnes, W.~T., Christe, S., {et~al.} 2023, Frontiers in
  Astronomy and Space Sciences, 10

\bibitem[{To {et~al.}(2021)To, Long, Baker, Brooks, van {Driel-Gesztelyi},
  Laming, \& Valori}]{toEvolutionPlasmaComposition2021}
To, A. S.~H., Long, D.~M., Baker, D., {et~al.} 2021, The Astrophysical Journal,
  911, 86, \dodoi{10.3847/1538-4357/abe85a}

\bibitem[{Tolbert \& Schwartz(2020)}]{tolbertOSPEXObjectSpectral2020}
Tolbert, K., \& Schwartz, R. 2020, Astrophysics Source Code Library,
  ascl:2007.018

\bibitem[{Tsuneta(1996)}]{tsunetaStructureDynamicsMagnetic1996}
Tsuneta, S. 1996, The Astrophysical Journal, 456, 840, \dodoi{10.1086/176701}

\bibitem[{Veronig {et~al.}(2005)Veronig, Brown, Dennis, Schwartz, Sui, \&
  Tolbert}]{veronigPhysicsNeupertEffect2005}
Veronig, A.~M., Brown, J.~C., Dennis, B.~R., {et~al.} 2005, The Astrophysical
  Journal, 621, 482, \dodoi{10.1086/427274}

\bibitem[{Warren(2014)}]{warrenMEASUREMENTSABSOLUTEABUNDANCES2014}
Warren, H.~P. 2014, The Astrophysical Journal Letters, 786, L2,
  \dodoi{10.1088/2041-8205/786/1/L2}

\bibitem[{Winebarger {et~al.}(2012)Winebarger, Warren, Schmelz, Cirtain,
  {Mulu-Moore}, Golub, \& Kobayashi}]{winebargerDEFININGBLINDSPOT2012}
Winebarger, A.~R., Warren, H.~P., Schmelz, J.~T., {et~al.} 2012, The
  Astrophysical Journal, 746, L17, \dodoi{10.1088/2041-8205/746/2/L17}

\end{thebibliography}

\appendix

\section{Supplemental material}

This section includes two static figures that display the flare eruptions analyzed in this study, with the associated animations available online.

\begin{figure}[ht!]%
	\centering
	\includegraphics[width=0.7\textwidth]{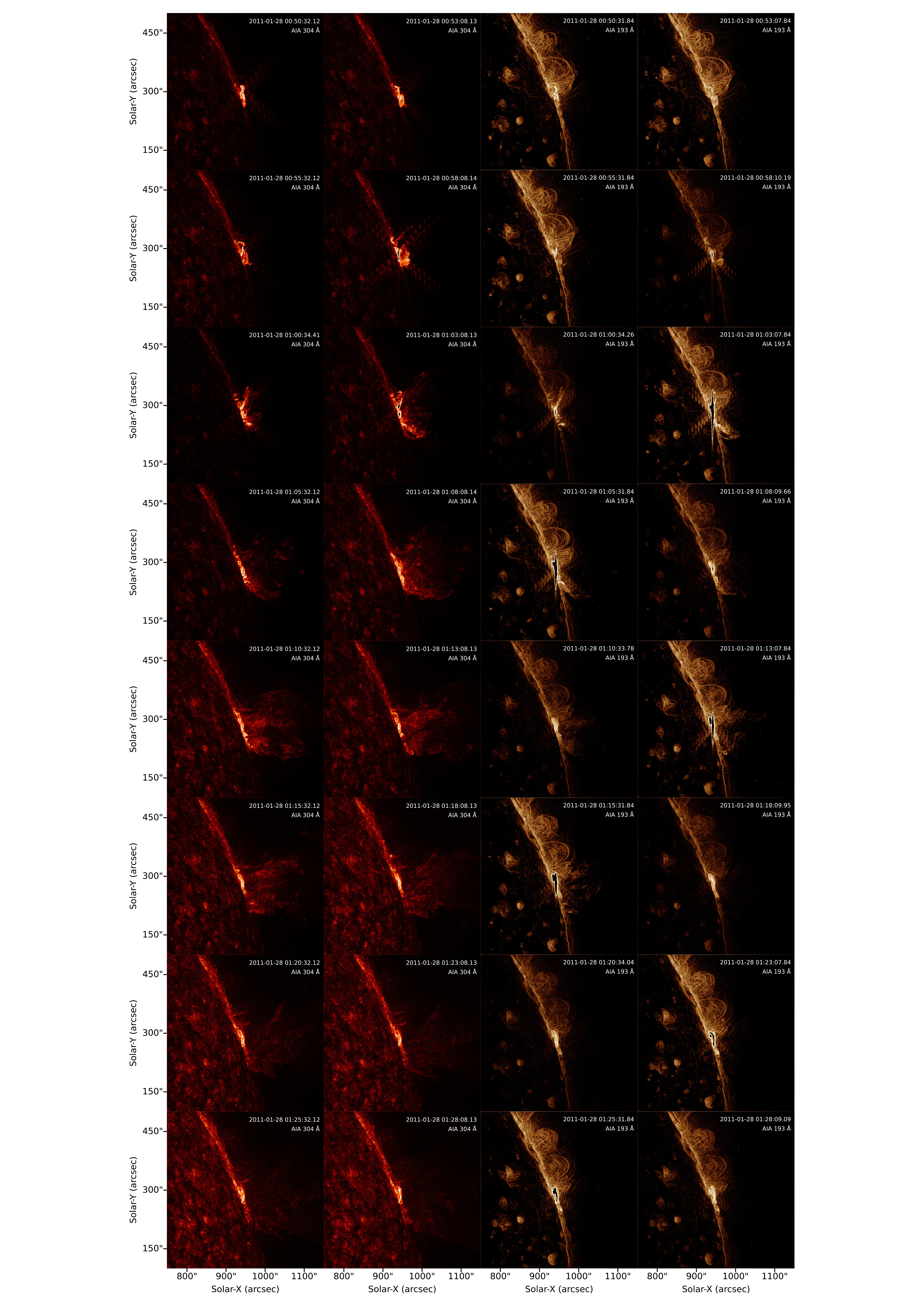}
	\caption{SDO/AIA 304~{\AA} (\ion{He}{2}, log$T$=4.7~K; left two columns) and 193~{\AA} (primarily \ion{Fe}{12}, log$T$=6.1~K; right two columns) images of flare FL1. The mosaic animation for both wavelengths is provided, with a duration of 7~seconds. The detailed description is provided in the main text (see Section~\ref{sec_eventoveriew}).}
	\label{fig_M1365AIA}
\end{figure}

\begin{figure}[ht!]%
	\centering
	\includegraphics[width=0.7\textwidth]{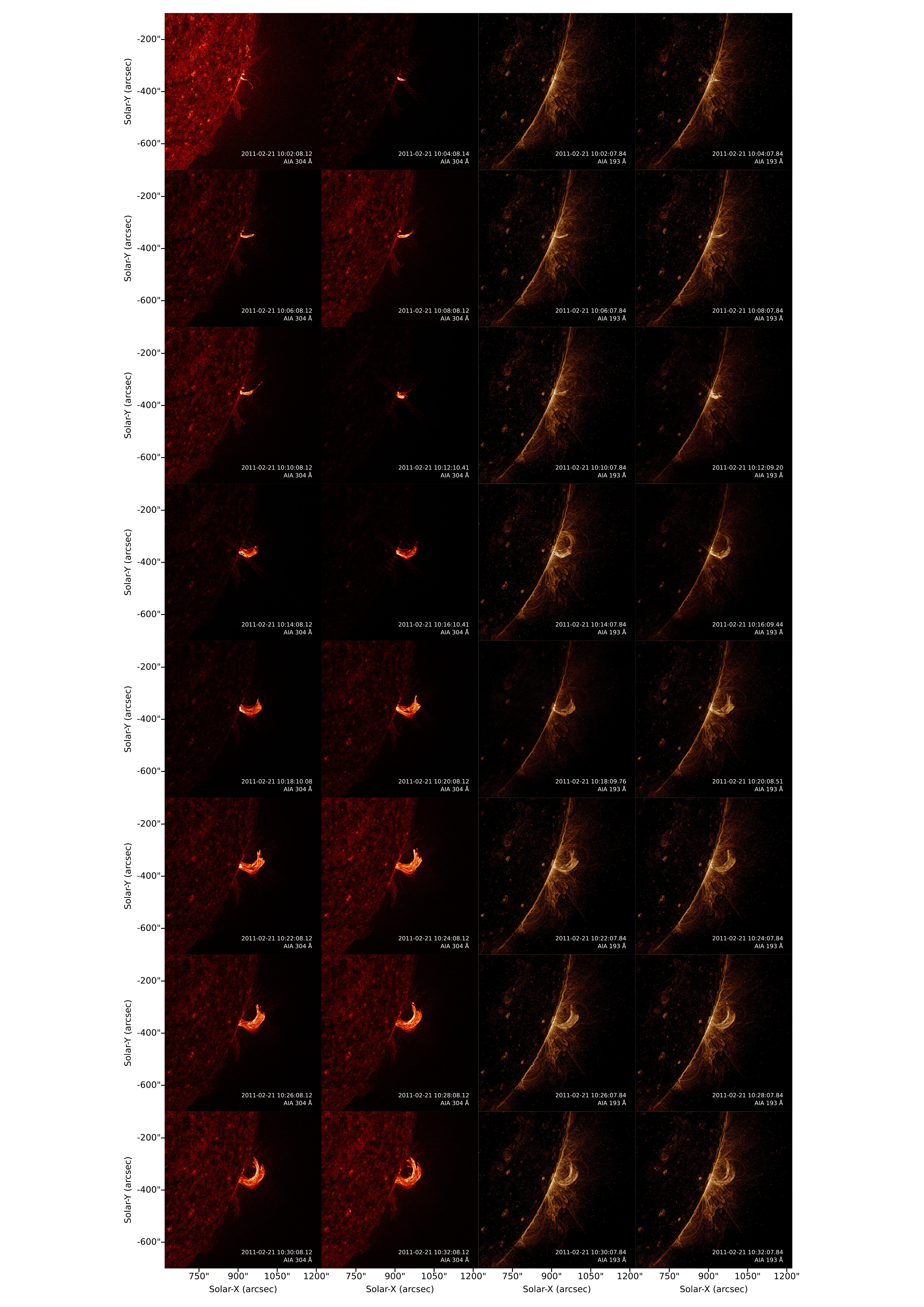}
	\caption{SDO/AIA 304~{\AA} (\ion{He}{2}, log$T$=4.7~K; left two columns) and 193~{\AA} (primarily \ion{Fe}{12}, log$T$=6.1~K; right two columns) images of flare FL2. The mosaic animation for both wavelengths is provided, with a duration of 6~seconds. The detailed description is provided in the main text (see Section~\ref{sec_eventoveriew}).}
	\label{fig_C1663AIA}
\end{figure}

\end{document}